\documentclass[times,twocolumn,final]{elsarticle}

\usepackage{cag}
\usepackage{framed,multirow}

\usepackage{amssymb}
\usepackage{latexsym}

\usepackage{url}
\usepackage{xcolor}
\definecolor{newcolor}{rgb}{.8,.349,.1}

\usepackage{hyperref}

\usepackage[switch,pagewise]{lineno} 

\usepackage{siunitx} 
\DeclareSIUnit{\lpi}{LPI} 
\DeclareSIUnit{\dpi}{DPI} 

\usepackage[inline]{enumitem} 

\usepackage{amsmath}
\newcommand\abs[1]{\left|#1\right|}

\usepackage{subcaption} 
\usepackage{algorithm}
\usepackage[noend]{algpseudocode}
\makeatletter
\def\BState{\State\hskip-\ALG@thistlm}
\makeatother

\usepackage{tikz}
\usetikzlibrary{positioning}

\makeatletter
\newcommand\footnoteref[1]{\protected@xdef\@thefnmark{\ref{#1}}\@footnotemark}
\makeatother

\journal{Computers \& Graphics}

\begin{document}

\verso{Preprint Submitted for review}

\begin{frontmatter}

\title{Hatching for 3D prints: line-based halftoning for dual extrusion fused deposition modeling}

\author[1,2]{Tim \snm{Kuipers}}
\author[2]{Willemijn \snm{Elkhuizen}}
\author[3]{Jouke \snm{Verlinden}}
\author[2]{Eugeni \snm{Doubrovski}}

\address[1]{Ultimaker, Watermolenweg 2, Geldermalsen, the Netherlands}
\address[2]{Delft University of Technology, Faculty of Industrial Design Engineering, Landbergstraat 15, Delft, the Netherlands}
\address[3]{University of Antwerp, Faculty of Design Sciences, Ambtmanstraat 1, Antwerpen, Belgium}

\received{April 21, 2018}
\accepted{April 30, 2018}
\availableonline{May 4, 2018}

\begin{abstract}
This work presents a halftoning technique to manufacture 3D objects with the appearance of continuous grayscale imagery for Fused Deposition Modeling (FDM) printers.
While droplet-based dithering is a common halftoning technique, this is not applicable to FDM printing, since FDM builds up objects by extruding material in semi-continuous paths.
The line-based halftoning principle called 'hatching' is applied to the line patterns naturally occuring in FDM prints, which are built up in a layer-by-layer fashion.
The proposed halftoning technique isn't limited by the challenges existing techniques face;
existing FDM coloring techniques greatly influence the surface geometry and deteriorate with surface slopes deviating from vertical
or greatly influence the basic parameters of the printing process and thereby the structural properties of the resulting product.
Furthermore, the proposed technique has little effect on printing time.
Experiments on a dual-nozzle FDM printer show promising results.
Future work is required to calibrate the perceived tone.
\end{abstract}


\end{frontmatter}


\begin{figure*}
	\centering
		\includegraphics[height=.7\columnwidth]{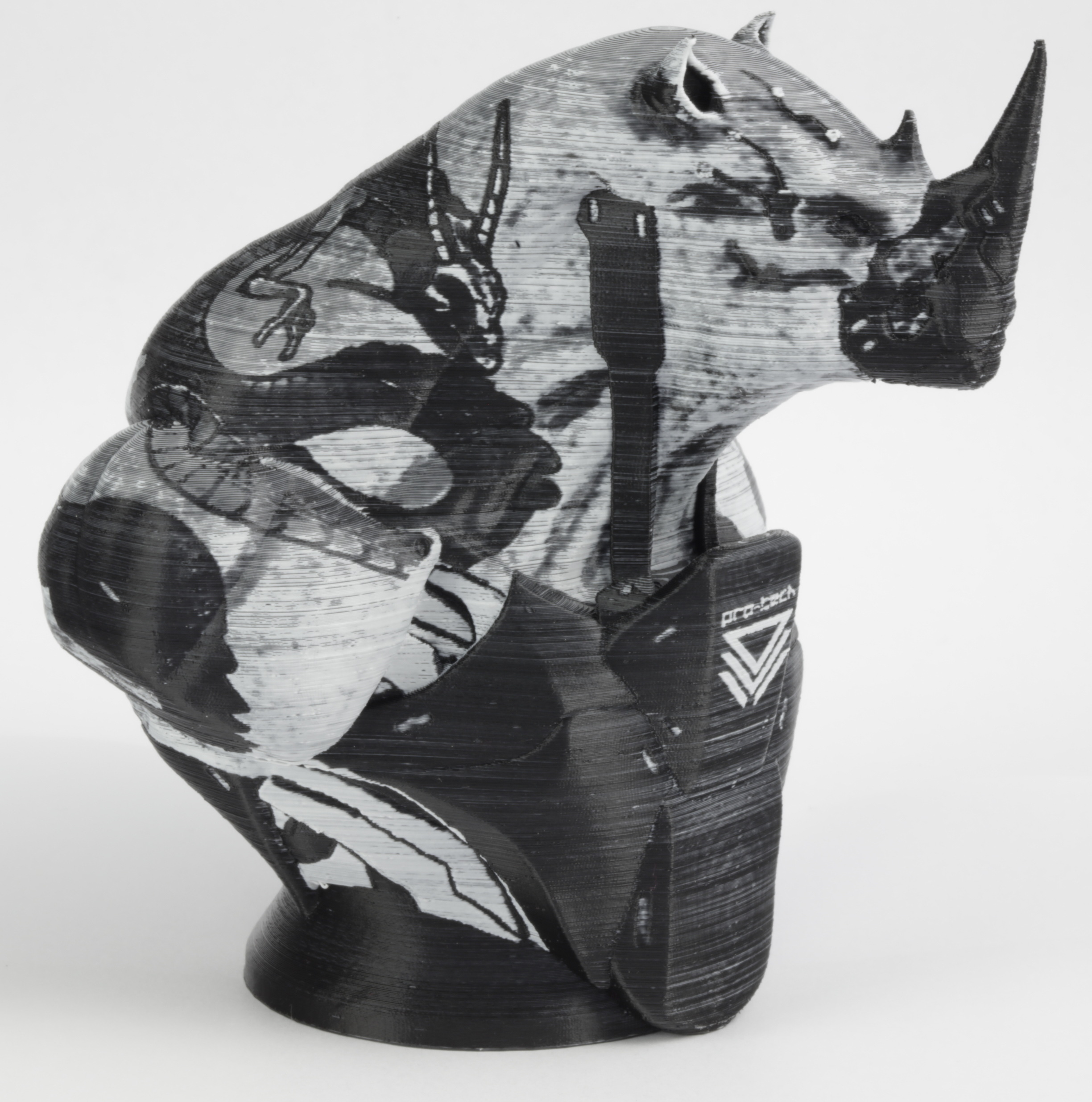}
		\includegraphics[height=.7\columnwidth]{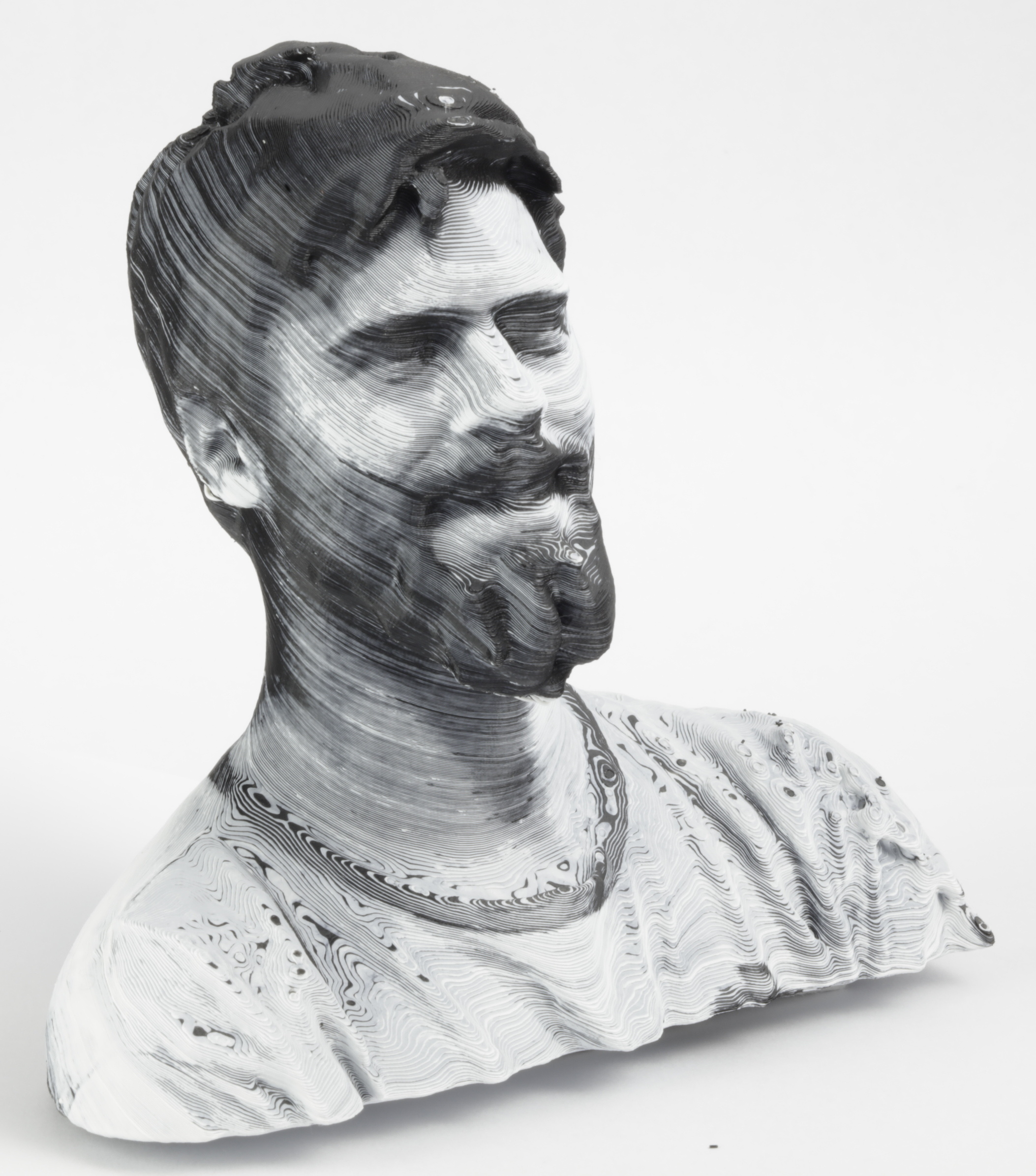}
		\includegraphics[height=.7\columnwidth]{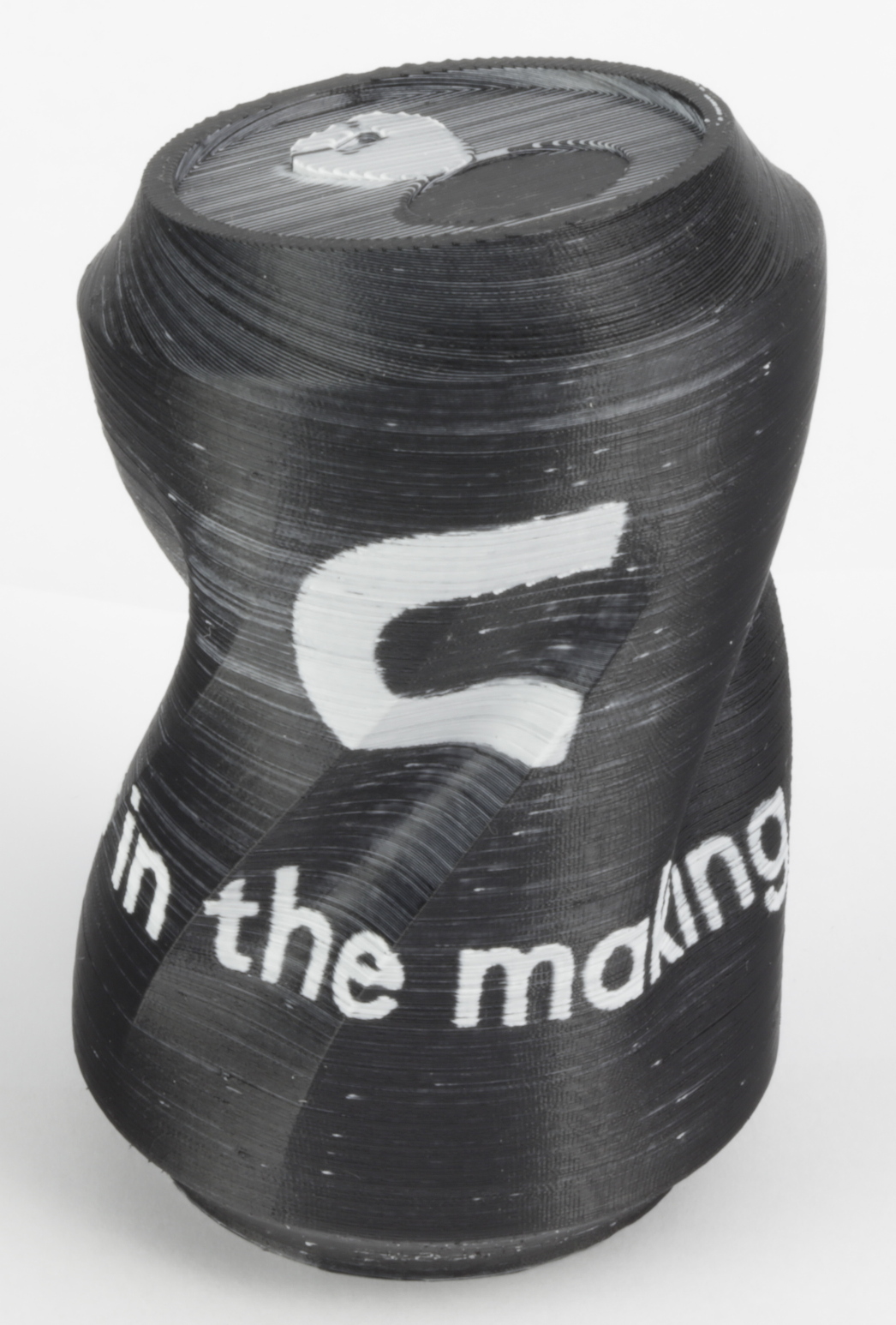}
		\includegraphics[height=.7\columnwidth]{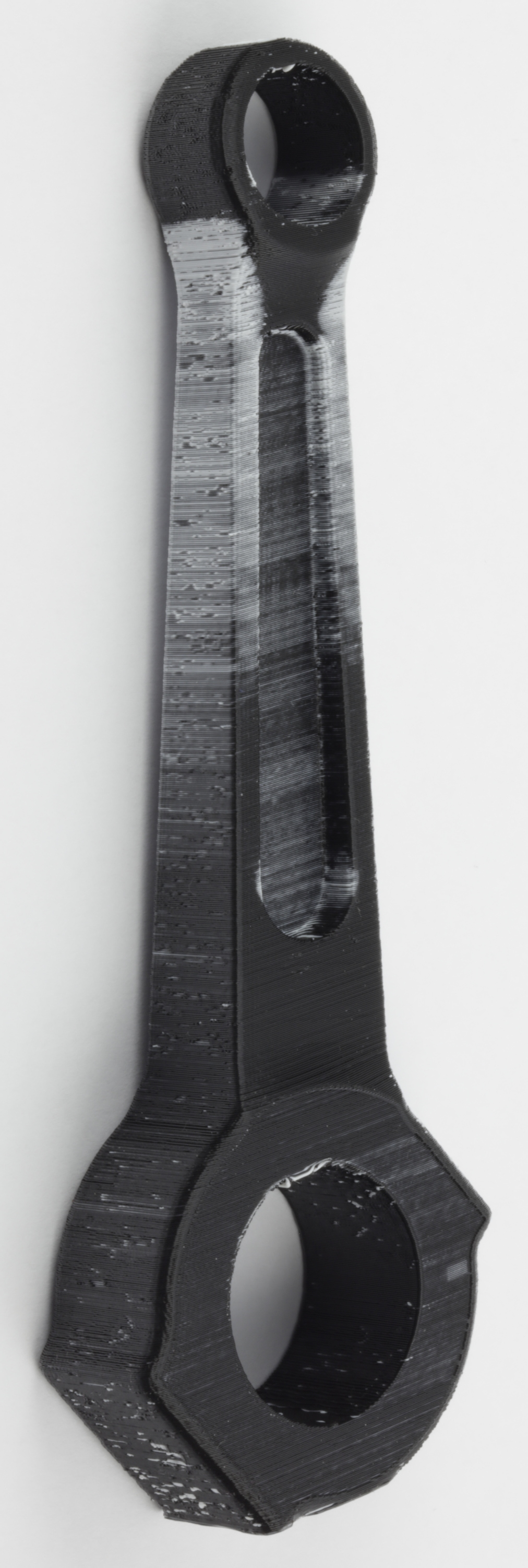}
	\caption{3D prints obtained by applying hatching on a 14 cm 3D portrait, a 15 cm artistic figurine, a full size soda can with textual information, and the result of a stress analysis performed on a 16 cm connecting rod of a piston engine.}
	\label{fig:teaser}
\end{figure*}

\section{Introduction}
The ability to apply color to 3D printed parts is relevant for both prototyping and manufacturing. 
Possible applications include reproduction of color-scanned 3D objects and fabrication of products with logos and labeling. 
Color can also be used as a design feature or to visualize geometric information such as the results of finite element analyses. 
See figure~\ref{fig:teaser}.

At present, 3D printing in color is available for a variety of Additive Manufacturing (AM) systems that are predominantly based on ink-jet technology.  
Techniques for printing color using only Fused Deposition Modeling (FDM) are sparse and suffer from a low resolution or have radical impact on the printing process.

This paper presents a novel technique for fabricating 3D grayscale objects using the FDM 3D printing method.
It uses a principle which is based on modulating the visible width of printed lines of two alternating colors to produce the appearance of continuous tone gradients. 
Creating the perception of continuous tones by generating small patterns of discrete colors is termed \emph{halftoning}. 

Implementing color variation with high frequency details using FDM is a challenge. 
FDM builds up objects by extruding material in semi-continuous paths, which makes it impossible to apply droplet-based halftoning principles that are commonplace in existing color 3D printers. 

A promising technique to fabricate continuous tone objects using FDM has been presented by \citeauthor{reiner2014dual}.
However, since that technique inherently produces textures at a relatively low sample rate, it does not allow the fabrication of high frequency details.
Furthermore, the technique does not allow fabrication of textures on horizontal surfaces and degrades for diagonal surfaces with a slope approaching horizontal.   

Addressing these issues, we propose a novel halftoning technique for dual-extrusion FDM systems.
The proposed halftoning technique is based on hatching, an established 2D halftoning principle based on lines rather than dots.
The implementation of the technique described in this paper is open source and can be found at \url{github.com/Ultimaker/CuraEngine} \cite{cura}.

This paper is an extension of the techniques proposed by conference paper by \citet{kuipers20173d}.
The technique proposed there changes the geometry of alternating black and white layers to modulate the perceived grayscale tone when viewed either from directly above or when viewed straight from the side.

Rather than presenting different hatching techniques for two viewing angles, this paper presents a unified hatching technique for viewing surfaces from any angle and in particular a viewing angle locally perpendicular to the surface.
The phenomenon that an overhanging line occludes the previous layer, also known as sagging, is exploited for an edge case of that general hatching technique.
This paper provides a model of the sagging behavior which is used to derive the proportion of visible white to black filament from any viewing angle.
Experimental data of sagging is collected and analysed in order to grasp the limitations of our model.

\section{Background}

\subsection{Process Planning for FDM}
The following section will briefly explain the basics of process planning for FDM, a.k.a. \emph{slicing}.
Some elementary concepts and processes are explained required for understanding the presented hatching technique.
The terminology employed and the techniques described apply to the open source slicing software called Cura\cite{cura}.

FDM generally builds up 3D prints in a layer-by-layer fashion.
One of the first stages in slicing is generating the outlines of each layer.
The outlines are the boundaries of the regions which are to be filled with material.
Line segments are generated by intersecting each triangle of the input mesh with horizontal planes at heights corresponding to each layer.
All line segments of a layer are then stitched into polygons which form the outlines of that layer.

\begin{figure}
      \centering
      \includegraphics[height=.5\columnwidth]{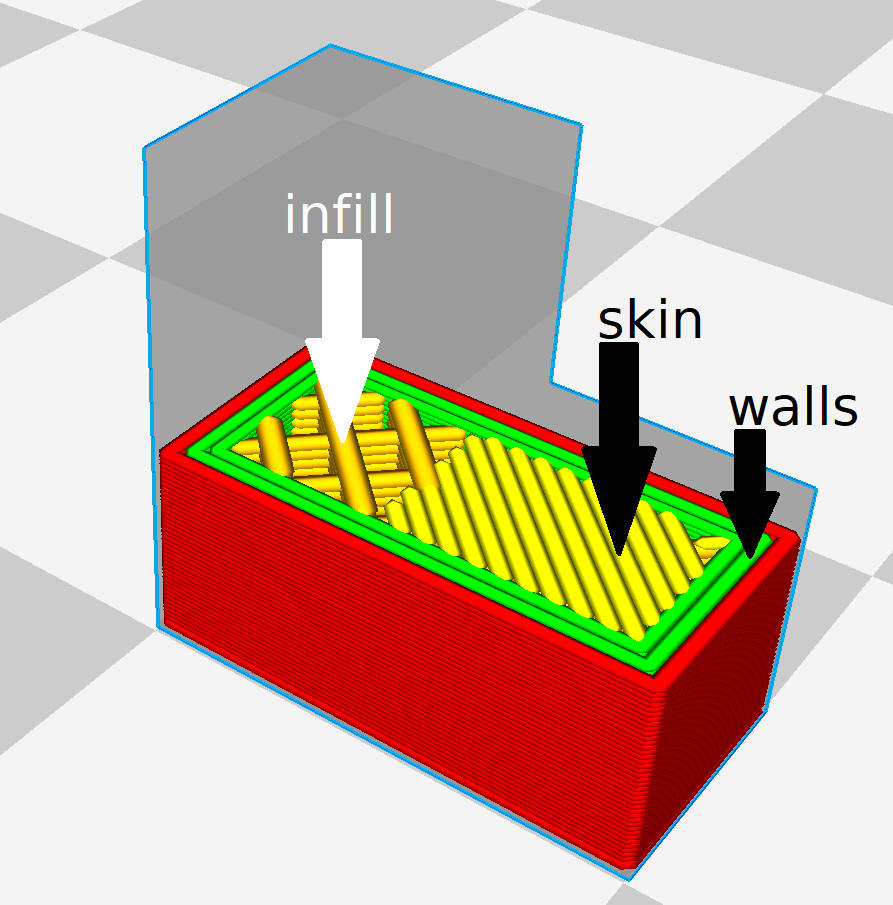}
      \caption{The lower part of an FDM process plan for an L-shaped 3D model. Several types of print path are indicated.}
      \label{fig:walls_skin}
\end{figure}

Because starting and stopping extrusion of filament causes blemishes, the outlines of a layer are achieved by following the contours of the layer: the walls.
Several consecutive walls are printed next to each other.
The outer walls are generated by applying an inward offset of half the line width to the outlines.
Successive walls are then generated by applying offsets to previous walls.
These walls define printed lines which follow the contours of the object.
See figure~\ref{fig:walls_skin}.

The remaining area within the innermost wall is split into infill and skin.
By applying boolean operations on the leftover region with the outlines of layers above and below we calculate the areas which are close to the top and bottom of the model boundary surface: the skin.
By applying a difference operation we can then determine the infill areas from the skin areas and the region left over from the walls.
The skins are several layers thick and they are densely filled with a pattern of parallel lines.


\subsection{Commercial color 3D printers}
The first commercial full color 3D printing systems date back to 1993 \cite{zcorp:2005}.
These systems use ink-jet technology to apply colored binder onto white powder \cite{iliescu2009z}. 
Consecutive layers of bound powder form the final 3D model.
Instead of jetting a binder onto a substrate, Mcor developed a process in which conventional ink is jetted onto sheets of paper, which are then cut and stacked \cite{Mcor2013}. 
Stratasys uses ink-jet technology to print the building material itself.
Their recent system incorporates six heads, each able to print a colored material \cite{Stratasys}.
More recently, HP Inc.\ introduced a printing technology in which liquid agents are jetted onto powder in order to alter the powder's fusing behavior. 
According to the company, these agents may also include color in the future \cite{HP}.

\subsection{3D Halftoning}
Because printers work with a limited number of base colors, specific strategies need to be applied to make full color prints.
In 2D printing, this is usually done through different halftoning techniques.

2D ink-jet technologies apply a halftoning principle called dithering. 
In dithering, the distance between printed colored dots is varied to create perceived variations of colors.
While halftoning for 2D printing industry is well developed, halftoning for 3D printers is still an active field of research. 
The first mention of halftoning in 3D printing is not focused on color reproduction but on material density variation for stereolithography \cite{Lou1998}.  
Techniques for 3D color dithering have been presented for binder jetting printers \cite{Cho2003851}.
\citeauthor{Vidim2013} presented a programmable pipeline for multi-material 3D printing. In their pipeline, dithering is applied for both visual and mechanical properties \cite{Vidim2013}.
With the aim to produce full-color prints using material jetting technology, \citeauthor{Brunton2015} presented a halftone technique that takes into account the translucency of the printed material. 

The different commercial systems discussed above all use ink-jet technology, and the color halftoning techniques considered are based on the ability of ink-jet to deposit discrete droplets of color in the micrometer range. 
This allows the production of high frequency details. 
However, FDM lacks the ability to deposit discrete features in this size range, since it builds up objects by extruding semi-continuous lines of material. 
Therefore, the halftoning techniques discussed above cannot be directly adapted to FDM. 
To create high frequency details using FDM, the development of novel halftoning techniques is needed.

\subsection{Color FDM}
Limited by the semi-continuous material extrusion of FDM, techniques have been proposed that aim to reproduce continuous tones in FDM prints.
These can be categorized into two tactics.
\begin{enumerate*}
  \item Mixing material and color prior to extrusion.
  \item Applying halftone principles.
\end{enumerate*}

\citeauthor{corbett2012reprap}'s FDM grayscale printing technique performs continuous color mixing by implementing a mixing nozzle in the print head \cite{corbett2012reprap}. 
In this setup, multiple feeders are connected to a single nozzle.
Multiple materials are fed into a volume of the nozzle where they are molten and mixed.
The main limitation of this technique is that color changes require the full volume inside the mixing nozzle to be flushed;
the horizontal resolution is limited by the length a printed line needs to be to flush the nozzle.

\citeauthor{reiner2014dual} have shown that a dot-based halftoning principle can be applied to dual-nozzle FDM systems in order to produce two-tone texture mapped 3D prints \cite{reiner2014dual}.
Their technique involves applying sine patterns to the outer contour of each layer, with alternating each layer between black and white filament and shifting the sine patterns by half the wave length every two layers.
A texture-based amplitude modulation is then applied in order to make the peaks of the one filament protrude more than the other, which results in a shift in perceived color toward the former filament.
The main challenge of this technique is to align the phase of the sine pattern across consecutive layers with outlines of a different geometry.
The horizontal resolution of the produced textures is limited by the sine's wavelength, which in turn is limited by the width of the extruded lines.
The application of the sine pattern therefore results in loss of high frequency details in the geometry.
Moreover, this technique works best for vertical surfaces and performs significantly less on surfaces with lower slopes. 

More recently \citet{Song2017} have demonstrated a continuous tone imagery technique for FDM printing.
Though they make use of a mixing nozzle to alleviate several calibration issues, the fundament of their technique is inherently applicable to any multi-extrusion system which supports two or more base materials.
Each layer of the 3D print consists of several sublayers of slightly translucent material with different color.
By varying the thickness of these sublayers the perceived color can take any color within the gamut spanned by the colors of the base materials.
The technique is limited by the smallest sublayer thickness which can reliably be achieved on an FDM printer.

While the technique is able to produce stunning results, relying on variable sublayer thickness can cause several problems.
Because the layer thickness is one of the most elementary properties of the printing process, many parameters may depend on it:
the overhang angle for determining where to place support structures, the required amount of infill, the optimal movement speeds during extrusion, the optimal cooling speed of the fans, etc. 
While it is reasonable to find a local optimum in print settings given a static layer thickness, it is hard to find out how all settings relate to each other in order to support varying layer thickness with optimal print settings.
A suboptimal combination of these settings might lead to structural problems such as a decreased tensile strength.

Another problem with varying layer thickness is that the nozzle may collide with higher regions of a layer when printing lower regions.
Most commercial nozzles have a flat horizontal ring around the hole from which material protrudes.
This flat part helps to fuse the printed line together with neighboring lines of the current layer.
When trying to print a line with a low thickness adjacent to a line with a high thickness, the nozzle collides with the already printed line, which can cause the print to fail.

\begin{figure}
      \centering
       \begin{subfigure}[t]{0.49\columnwidth}

          \centering
          \includegraphics[width=.73\columnwidth]{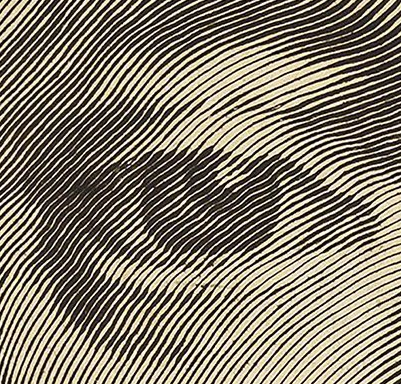}
          \caption{Closeup of `The Sudarium' \citep{mellan}.
            \footnote{
              Image is in the public domain.
            }   
          }
          \label{fig:claude_mellan}
      \end{subfigure}
      \begin{subfigure}[t]{0.4\columnwidth}
          \centering
          \includegraphics[width=.85\columnwidth]{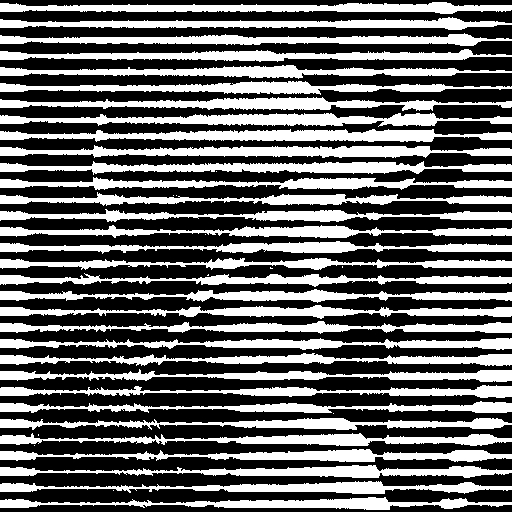}
          \caption{Engraving-style halftone image.}
          \label{fig:engraving_style}
      \end{subfigure}
      \caption{Examples of 2D hatching.}
\end{figure}

\subsection{Hatching}
Hatching is a halftoning principle which has lines as elements, rather than dots, which are commonplace in halftoning techniques.
Hatching dates back to  17\textsuperscript{th} century engraving techniques \cite{lavin2004claude}.
An example of this can be seen in figure~\ref{fig:claude_mellan}.
Variations in the perceived tone are achieved by varying the local ratio between the width of black lines and the width of the white background surrounding it \cite{praun2001real}.
In more recent developments, image processing algorithms have been proposed which convert grayscale images into black and white engraving-style halftone images \cite{yamamoto2006producing}. 
Their results resemble figure~\ref{fig:engraving_style}, which was produced using the linear Newsprint filter from the GNU Image Manipulation Program. 
The linear characteristic of hatching makes it a particularly suitable technique for halftoning in FDM.


\section{Method}
\begin{figure}
    \centering
\iftrue
  	 \begin{subfigure}[b]{0.45\columnwidth}
        \centering
         \includegraphics[width=\columnwidth]{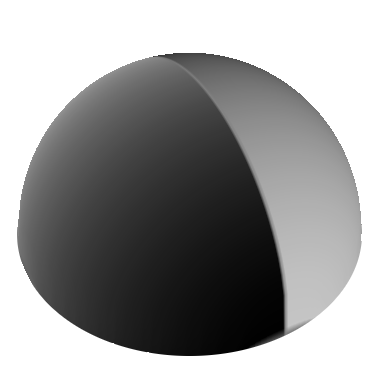}
        \caption{Model}
    \end{subfigure}
    \begin{subfigure}[b]{0.45\columnwidth}
        \centering
         \includegraphics[width=\columnwidth]{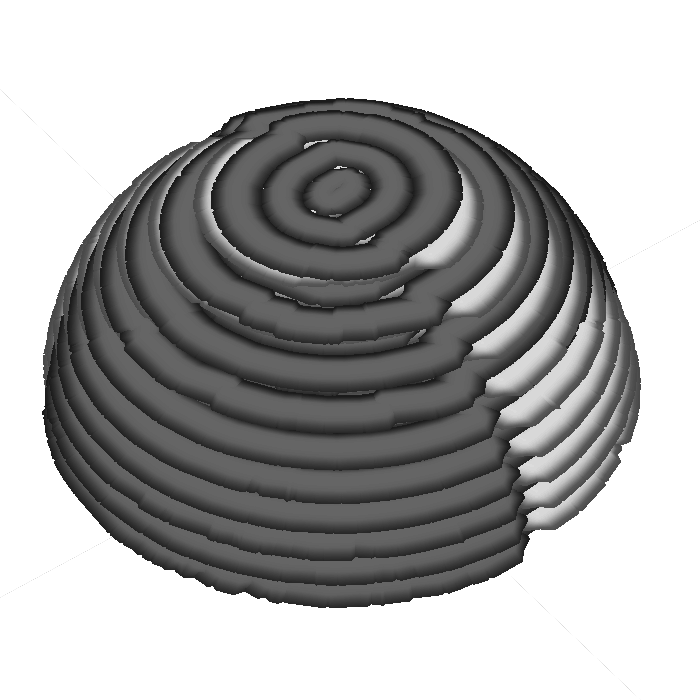} 
          \caption{Toolpaths}
    \end{subfigure}
\fi
    \caption{
        Example of a \SI{4.5}{\milli\meter} textured half dome and a visualization of the corresponding toolpaths.
    }
    \label{fig:example}
\end{figure}

The presented halftoning technique leverages the discrete nature of FDM.
Layers are printed with alternating filament.
All even layers are printed in black while all odd layers are printed in white.
The region of a layer which is visible from the outside resembles a line.
Changing the perceived widths of these lines controls the observed grayscale tone.
This is done by changing the outlines of a layer.
See figure~\ref{fig:example}.

At places where the model should be darker the outlines of the black layers are expanded and the outlines of the white layers are contracted, so that more black filament will be visible on the surface.
Because only the outlines are changed while the patterns with which these outlines are filled up remain the same,
the structural properties of 3D prints remain unaltered.

At places on the model where the texture tone is moderate or where the surface slope is high,
achieving the desired ratio of perceived material simply follows from the viewing angle.
For extremal texture tones or near vertical surfaces a more elaborate approach is needed, which is based on sagging.
In this section we first illustrate the simple hatching method after which the method is expanded to also make use of sagging.

\subsection{Hatching}

Because objects are built up by discrete planar layers, sloped surfaces are discretized;
this effect is knows as the stair-stepping effect.
By applying varying offsets to the outline polygons of each layer we can modulate the local widths of these stair steps.
A local grayscale tone arises from the proportion of white filament visible w.r.t.\ black filament at that region.
In places where the model needs to be light according to the surface texture color information of the 3D model, the white layers are offset outward and the black layers are offset inward.

The amplitudes of the variable offset should depend on the basic characteristics of the stair-stepping effect.
In regions where the surface is nearly horizontal the width of the stair steps is large and so those regions require a larger offset to change the ratio between visible white and black filament.

The amplitudes of the offset should also depend on the viewing angle.
When viewing an object from a higher altitude, the sides of the steps have less impact then when viewing the stair steps from a lower altitude.

\begin{figure}
    \centering
    \begin{subfigure}[t]{0.85\columnwidth}
        \centering
        \includegraphics[width=\columnwidth]{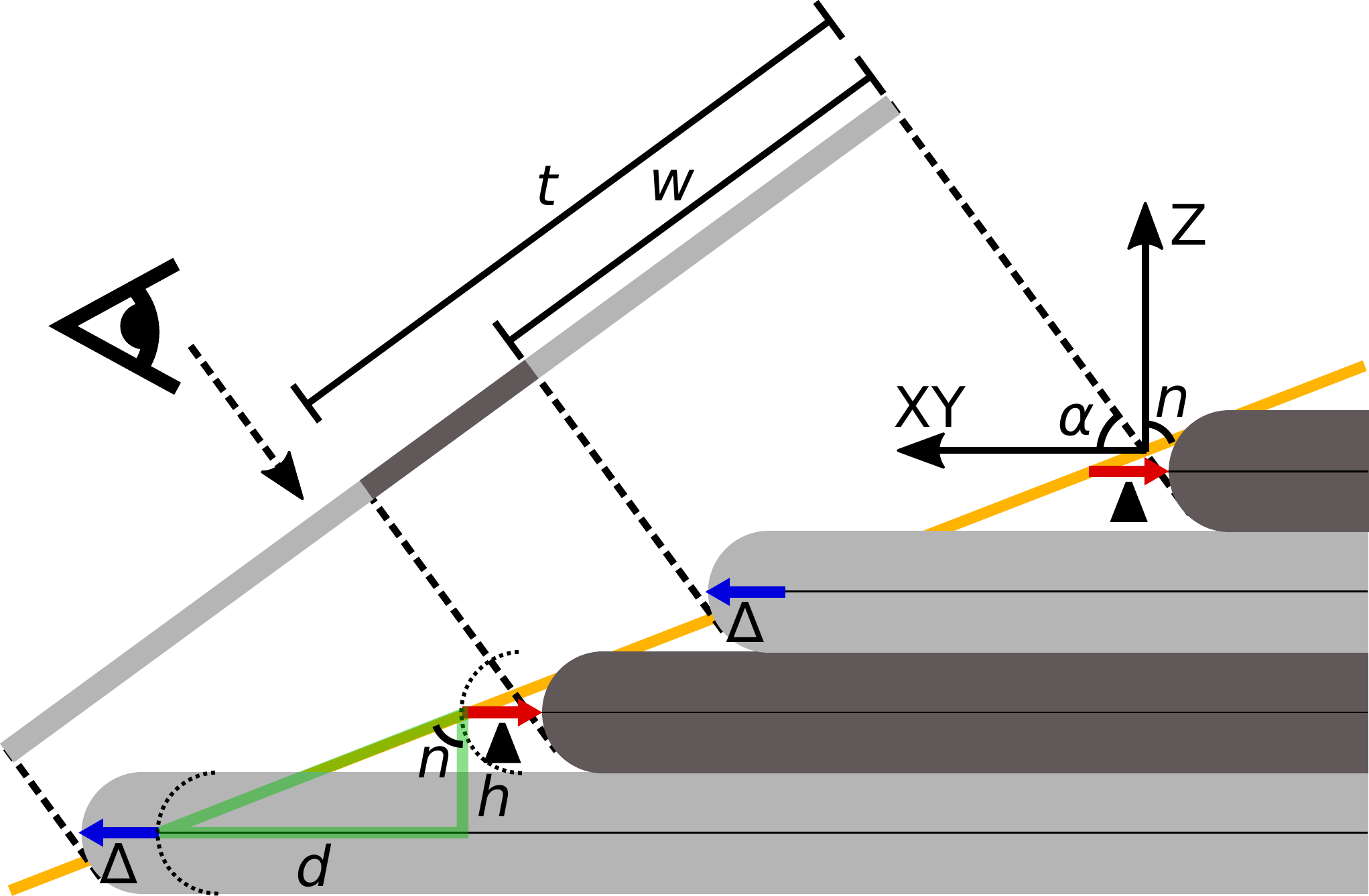}
    \end{subfigure}
    \caption{
        Vertical cross-section of a print showing the basic hatching technique.
        The cross-section is perpendicular to the local surface and to the build plate.
        The model surface is indicated in orange.
        The applied offsets in red and blue cause the amount of visible dark filament to be less than the amount of visible light filament.
    }
    \label{fig:hatching}
\end{figure}

The ratio $r$ of visible white to the total of visible white and black can then be calculated by projecting the horizontal component (the width of the stair step) and vertical component (the layer height) onto the viewing plane.
The perceived ratio $r$ and the corresponding offset $\Delta$ are therefore related by the following formula:

\begin{align}
\Delta &= -\blacktriangle      \nonumber
\\
r &=  \frac{w}{t} = \frac{  \left(   d + 2 \Delta  \right) \sin(\alpha)     + h \cos(\alpha)     }{  2d \sin(\alpha)  + 2h\cos(\alpha)  } 
  \label{eq:hatching_luminance}
\\
d &= h \tan(n)      \nonumber
\end{align}

, where the ratio $r$ is a value between $0$ and $1$,
$\blacktriangle$ is the offset applied to black layers and $\Delta$ to white,
$w$ is the projection of visible white filament onto the viewing plane,
$t$ is the projection of two full stair steps into the viewing plane,
$0 \leq n \leq 0.5 \pi$ is the absolute angle between the surface normal vector $\vec{n}$ and it's projection downward on the horizontal plane,
$\alpha$ is the absolute viewing angle, 
$h$ is the layer thickness
and $d$ is the horizontal stair step distance corresponding to the given surface slope.
See figure~\ref{fig:hatching}.

From the texture image we can derive the required ratio of visible white filament $r$.
Disregarding lighting conditions, the amount of visible white filament at a given location should be proportional to the luminance of the corresponding texture image coordinates.
(Section~\ref{section:texture_mapping} covers mapping outline locations to texture UV coordinates.) 
The luminance values are obtained by applying a gamma expansion to the Luma component of the pixel: $r = (0.2126R+0.7152G+0.0722B)^{1/2.2}$ \cite{poynton2004color}.
Using formula~\ref{eq:hatching_luminance}, we can then obtain a formula for the required offset at a given location on the model:

\begin{align}
\Delta     &=   h \left(    \frac12  +   r   \right)      \frac{      \sin(n)  \sin(\alpha)  + \cos(n)\cos(\alpha)      }{     \cos(n)  \sin(\alpha)    }
  \label{eq:hatching_delta}
\end{align}

Performing trigonometric functions on a computer is a relatively expensive task.
Luckily, given that $0 \leq n \leq 0.5 \pi$, they can be derived from the $x$, $y$ an $z$ component of the vector: $\sin(n) = \abs{ \vec{n}_z  }$ and $\cos(n) =  \sqrt[]{\vec{n}_x^2 + \vec{n}_y^2}$.

The formula above calculates the offset required to optimize the perceived luminance value \emph{w.r.t.\ a certain viewing angle}.
For other viewing angles the perceived luminance value will be off.
Choosing a viewing angle could be a user input which depends on the model and its function.

An alternative is to optimize the perceived tone w.r.t.\ a viewing angle perpendicular to the surface.
In such a case  the normal angle $n$ aligns with the viewing angle $\alpha$ and the above formulae simplify to:

\begin{align}
r &= \frac12 + \Delta \frac{ \sin(n) \cos(n) }{h}
\\
\Delta &= h \frac{\frac12 + r} {\vec{n}_z  \text{ } \sqrt[]{\vec{n}_x^2 + \vec{n}_y^2}}
  \label{eq:hatching_simple_delta}
\end{align}

It should be noted that these formulae are independent of the geometry of a cross-section of a layer.
As long as the cross section of a white layer follows the same geometry as a black layer, it doesn't matter whether the side of a layer is straight, a semicircle or something in between.


\begin{figure}
        \centering
    \begin{minipage}[t]{.4\columnwidth} \centering \includegraphics[height=1.5\columnwidth]{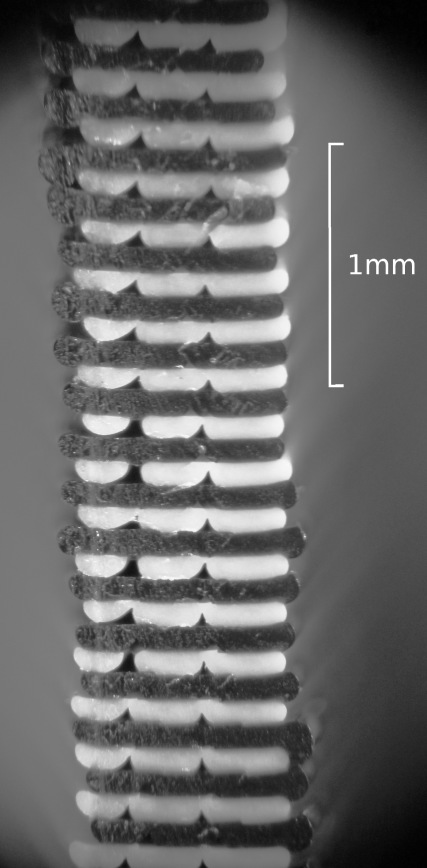} 

\smallskip

 \includegraphics[height=0.9\columnwidth]{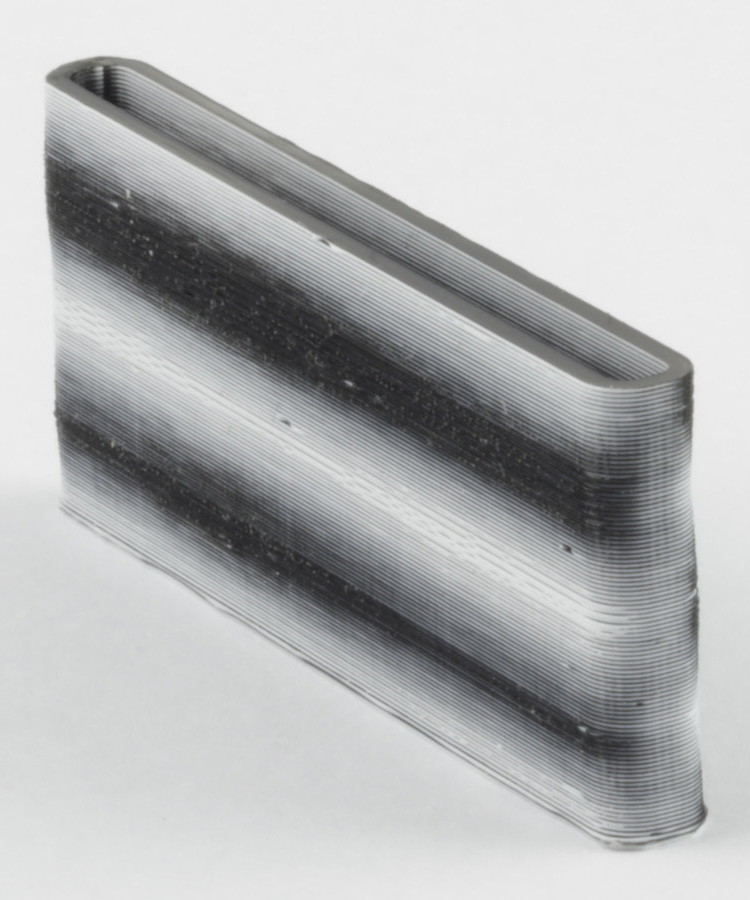} \subcaption{Vertical surface.} \end{minipage}
    \begin{minipage}[t]{.4\columnwidth} \centering \includegraphics[height=1.5\columnwidth]{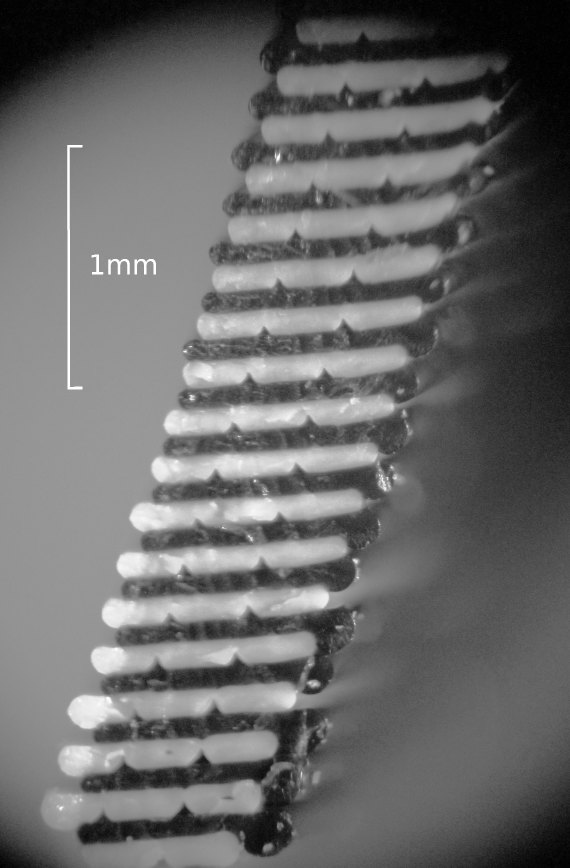}

\smallskip

 \includegraphics[height=0.9\columnwidth]{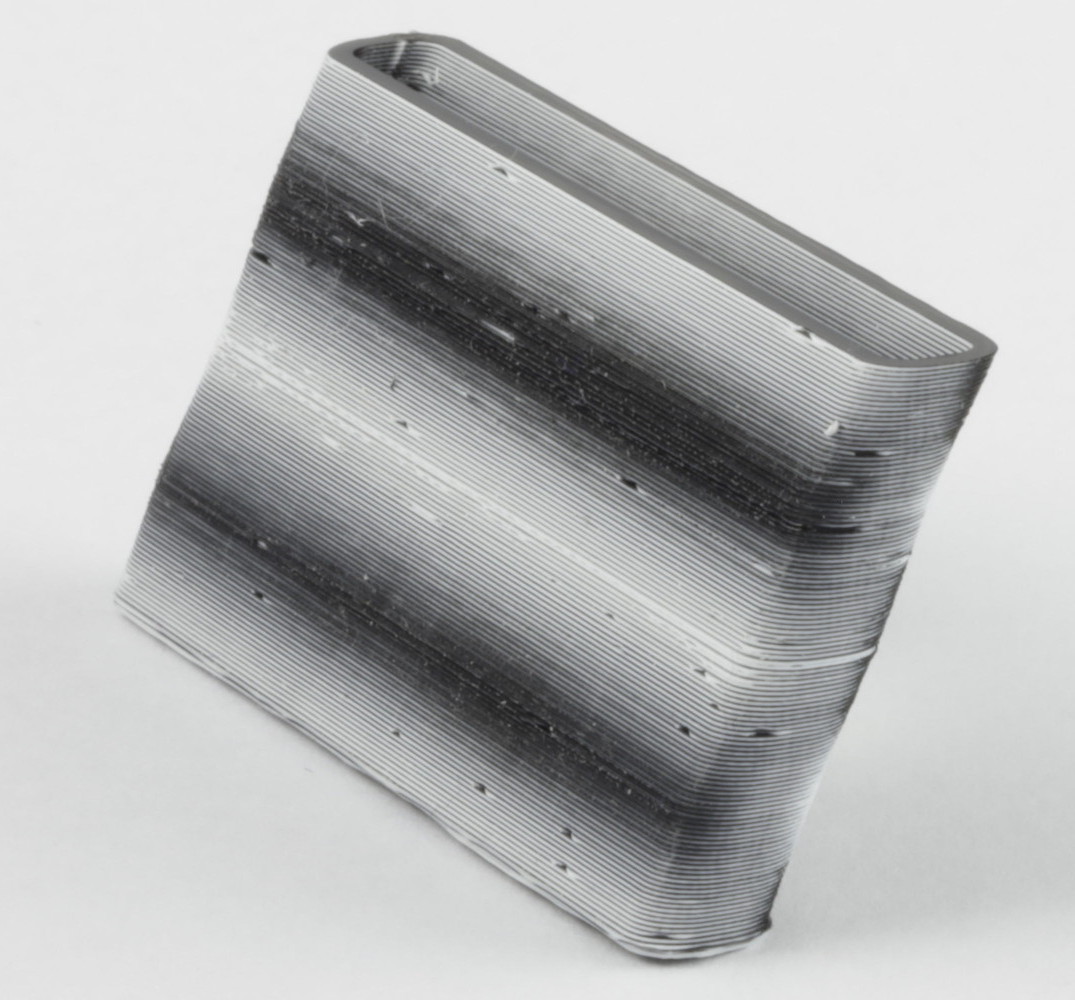} \subcaption{Sloped surface.}  \end{minipage}
    \caption{
    	Cross-section photos of sagging made with the Olympus SZ61 microscope.
        The specimen have various overhang distances while the layer thickness is \SI{0.1}{\milli\meter}.
    	The grayscale levels have been adjusted to increase the contrast in the dark and white regions at the cost of contrast of in between tones, so that details in both black and white filament are clearly visible.
    }  
    \label{fig:micro}  
\end{figure}

\subsection{Sagging}

However, when a layer is offset by an amount such that it extends beyond the previous layer, i.e.\ when $2\abs{\Delta} > d$,  the geometry of the cross-section of a layer will change.
The phenomenon of an overhanging layer occluding the layer below is knows as \emph{sagging}.
Figure~\ref{fig:micro} shows cross section microscopy photos of a print with various different overhang distances.
\citeauthor{reiner2014dual} proposed that this effect could be used to perform halftoning in FDM:
``subtle geometric offsetting between layers enables the control of variation of tone due to occlusion and gravity.''
This section explores ways in which to accurately use sagging for hatching in FDM.

Though gravity could influence sagging, it is conceivably a phenomenon caused by the uneven distribution of back pressure due to there being material of the previous layer under only some locations beneath the nozzle.
When printing a line on top of a previous layer, the existing layer creates back pressure which forces the material from the round shape of the nozzle to a flatter line.
This back pressure drops where there is no layer below, so given the same amount of pressure in the nozzle, more material is extruded there.
Given that there is no reason for that material to stay in the plane of that layer it follows that it will start to occlude the layer below.

\begin{figure}[b]
    \centering 
    \begin{minipage}[t]{.49\columnwidth} \centering 
    	\includegraphics[height=.49\textwidth]{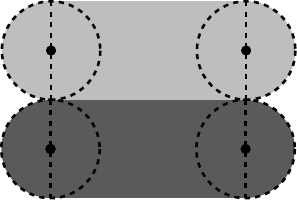} 
    	\subcaption{A printed line as a rectangle and two circles.} 
    \end{minipage}
    \begin{minipage}[t]{.49\columnwidth} \centering 
      \includegraphics[height=.49\textwidth]{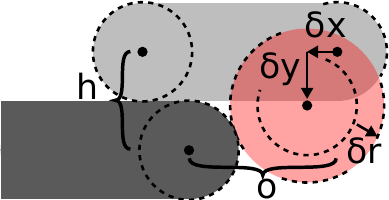} 
    	\subcaption{Sagging as a change in radius and position of the circle.}  
	\label{fig:sagging_model_overhang}
    \end{minipage}
    	\caption{Cross-sections of the geometrical model of a printed line.}  
      \label{fig:sagging_model}  
\end{figure}

\begin{figure*}
    \centering
    \begin{subfigure}{0.09\textwidth}
	\includegraphics[width=\textwidth]{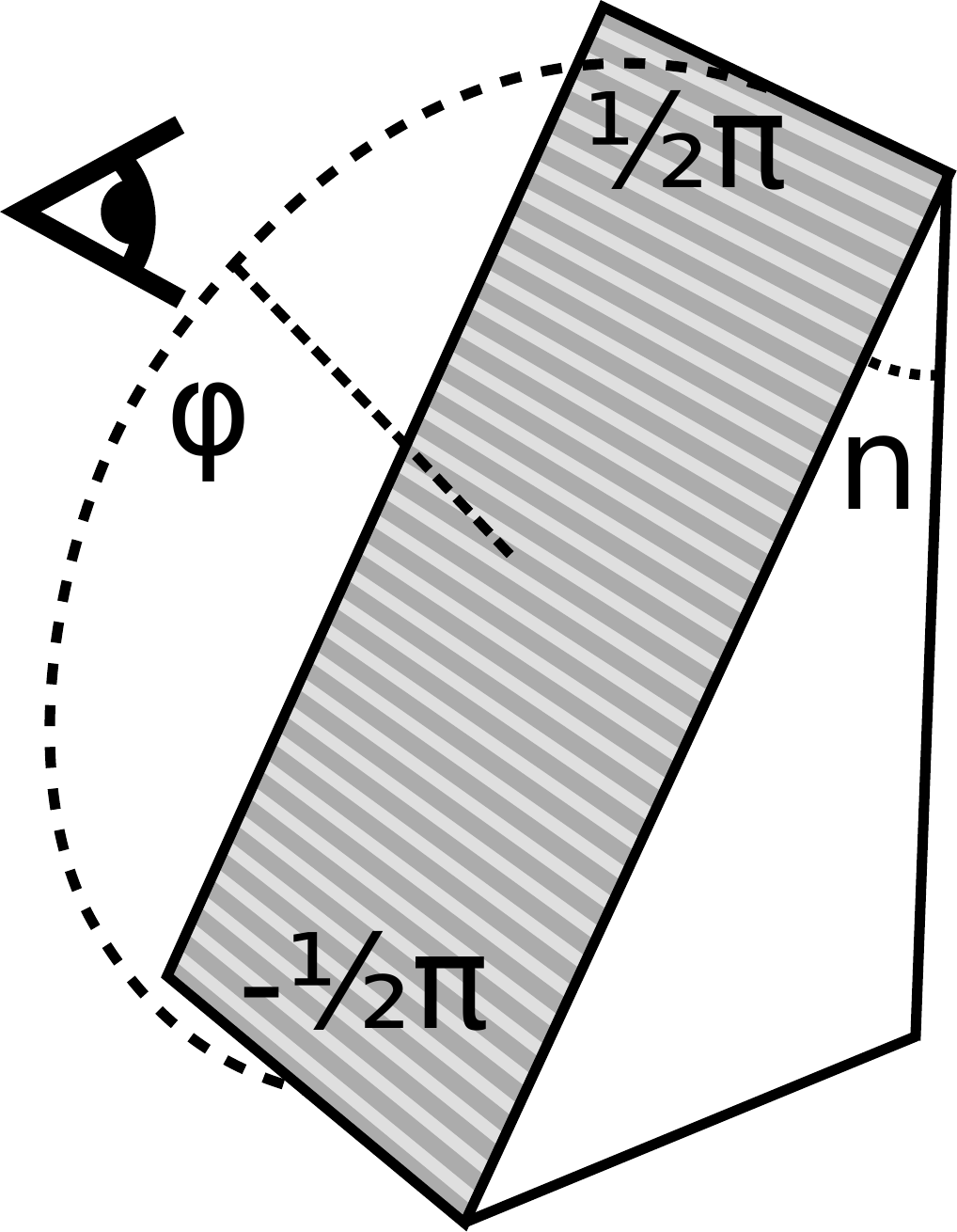}
	\label{fig:responde_viewing}
    \end{subfigure}
    \begin{subfigure}{0.9\textwidth}
	\includegraphics[width=\textwidth]{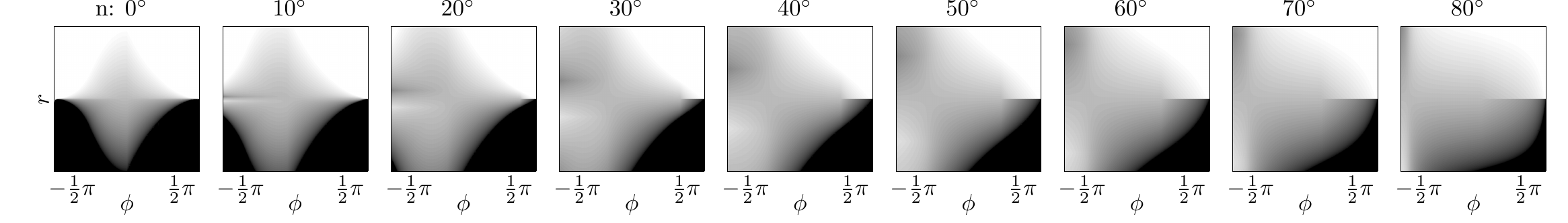}
	\label{fig:response_graphs}
    \end{subfigure}
    	\caption{
	Calculated luminance according to the geometrical model optimized for perpendicular viewing under various viewing angles and various ratios of visible white to total filament $r$.
	$\phi$ is the viewing angle w.r.t. the surface: $\phi = \alpha - n$.
	The image on the left shows how $\phi$ and $n$ relate to the print.
	}  
    	\label{fig:response_functions}  
\end{figure*}

\subsubsection{Modelling Sagging}
In order to produce the right grayscale tone we need to know how the occlusion relates to the amount of overhang.
The proportion of white to black filament visible from a given angle is influenced by the amount occlusion due to sagging.
A model which predicts the amount of visible filament from any viewing angle forms the basis of a grayscale tone calibration.

A geometrical model of the cross-section of a sagged line can provide a mathematical formula for calculating the amount of visible black and white filament from any viewing angle.
We can model the cross-section of a printed line as a rectangle and two circles for which the center and the radius depend on the amount of overhang.
See figure~\ref{fig:sagging_model}.
For simplicity we assume that the black and white filament exhibit the same sagging behavior.
The amount of occlusion in our model follows the following formula:


\begin{align}
f(o) &= (o  - \delta x) \sin(\alpha) + \delta y \cos(\alpha)+ \delta r 
\label{eq:sagging}
\end{align}

for positive angles $\alpha$
, where 
$o$ is the overhang distance by which the top line is extended beyond the bottom line,
$f$ is the amount of occlusion as viewed from $\alpha$ as a function of the amount of overhang
and
$\delta x$, $\delta y$ and $\delta r$ are the change in position and radius of the circle which is used to describe the side of tha sagged line.

Supplementing formula~\ref{eq:sagging} to formula~\ref{eq:hatching_luminance}, the perceived luminance in case of sagging when viewed from positive $\alpha$ becomes as follows:
\begin{align}
r_\neg &=    \frac{  \left(   d + 2 \Delta_\neg  \right) \sin(\alpha)     + h \cos(\alpha)  + s   }{  2d \sin(\alpha)  + 2h\cos(\alpha)  }  
\\
s &= 
f(\max(0, 2 \Delta - 2d)) - f(\max(0, 2 \blacktriangle - 2d))
\end{align}

If we specify that the circle of the sagging layer touches the layer below then one can easily see that $(o-\delta x)^2+(h-\delta y)^2 = (h+\delta r)^2$.
See figure~\ref{fig:sagging_model_overhang}.
When we further specify that the top of the circle coincides with the top of the layer $\delta y = \delta r $ and that the circle touches the previous layer, we can use the Pythagorean theorem to derive that $\delta r  = (o  - \delta x)^2 / 4h$.  
For simplicity we model the relation between the circle center receding and the overhang as a linear function going through the origin: $\delta x = Cx . o$.
Using the Pythagorean theorem we derive that  $ Cx = 1 - \sqrt[]{2} h / w$, where $w$ is the overhang distance at which a layer fully occludes the previous layer when viewed from the side.
From preliminary experiments we concluded that at $w = \SI{0.2}{\milli\meter}$ a layer of $h=\SI{0.1}{\milli\meter}$ would occlude the previous layer fully, which leads to $Cx \approx 0.2929$.

We can then use the quadratic formula to derive the offset required for a viewing angle perpendicular to the surface from the equations above, which results in algorithm~\ref{alg:offset}:

\begin{algorithm}[tb]
\begin{algorithmic}[1]
\Procedure{getOffset}{}
\State $n \gets getN()$\Comment{Normal vector of the model surface}
\State $L \gets getL()$\Comment{Luminance of the texture image}
\State $r \gets L^{1/\gamma}$\Comment{Gamma expansion}
\State $sin\_n \gets \abs{n.z}$
\State $cos\_n \gets \sqrt{n.x^2 + n.y^2}$
\State $tan\_n \gets sin\_n / cos\_n$
\State $diag \gets h / sin\_n$
\State $d \gets diag * cos\_n$
\State $dir \gets 1$
\If {$r > 0.5$}
\State $r \gets 1 - r$
\State $dir \gets -dir$
\EndIf
\State $\Delta \gets ((0.5 - r) / cos\_n) diag$
\If {$2 \abs{\Delta} \leq d$}
\State \textbf{return} $dir * \Delta$
\EndIf
\State $ Cx \gets getCx()$\Comment{Ratio of receding to overhang}
\State $ C \gets 1 - 2Cx + Cx^2$
\State $ a \gets -0.5 / (h * diag)  C  (1 + sin\_n)$
\State $ b \gets 0.5 / diag  (C * tan\_n  (1 + sin\_n) + 2  cos\_n  (Cx - 1))$
\State $ c \gets 0.5 - 0.5 cos\_n  (C/4 * tan\_n  (1+sin\_n) - Cx * cos\_n ) - r$
\State $ det \gets \max(0, b^2 - 4 a * c)$
\State $ \Delta \gets (-b - \sqrt{det} ) / (2 a)$
\State \textbf{return} $dir * \Delta$
\EndProcedure
\end{algorithmic}
\caption{Calculating offset distance}\label{alg:offset}
\end{algorithm}

Note that this algorithm only accurately describes our geometrical model for positive surface slopes, and also positive viewing angles because it is optimized for perpendicular viewing.
When relating the perceived proportion of white filament to the offset for viewing angles below zero one should take great care in distinguishing between different occlusion scenarios.
When the sagging layer is farther from the viewing plane than the layer below, the layer below occludes the bottom of the sagging layer.

Figure~\ref{fig:response_functions} plots the luminance as function of the viewing angle, surface angle and input grayscale tone.
It should be noted that the middle of these graphs - corresponding to a perpendicular viewing angle - coincides with the full gradient from black to white for each model angle.
Higher viewing angles result in higher contrast because of the parallax effect.
Extreme negative angles result in grayscale tones at the opposite side of the spectrum compared to the input tone;
this is due to the fact that sagging causes the top of the layer to be more round, which reduces the amount of visible filament from the sagged layer.

Negative surface slopes pose a plethora of problems.
Because there are viewing angles at which the sagging reduces the amount of visible filament, there can be combinations of surface normal angle and input tone which have multiple offsets as solution.
If the algorithm to compute the offset from the normal and the input luminance isn't stable, the perceived ratio of black to white filament can vary wildly for texture colors which only vary mildly.

Note that there is a maximum overhang up to which our sagging model makes sense.
If the overhang distance is larger than the line width, no more sagging can occur.
Moreover, at a surface slope of about \SI{-45}{\degree} the surface requires a support structure, which inevitably also affects the sagging behavior.


\subsection{Horizontal Hatching}
\begin{figure}
    \centering
    \begin{subfigure}[t]{0.25\columnwidth}
        \centering
        \includegraphics[width=\columnwidth]{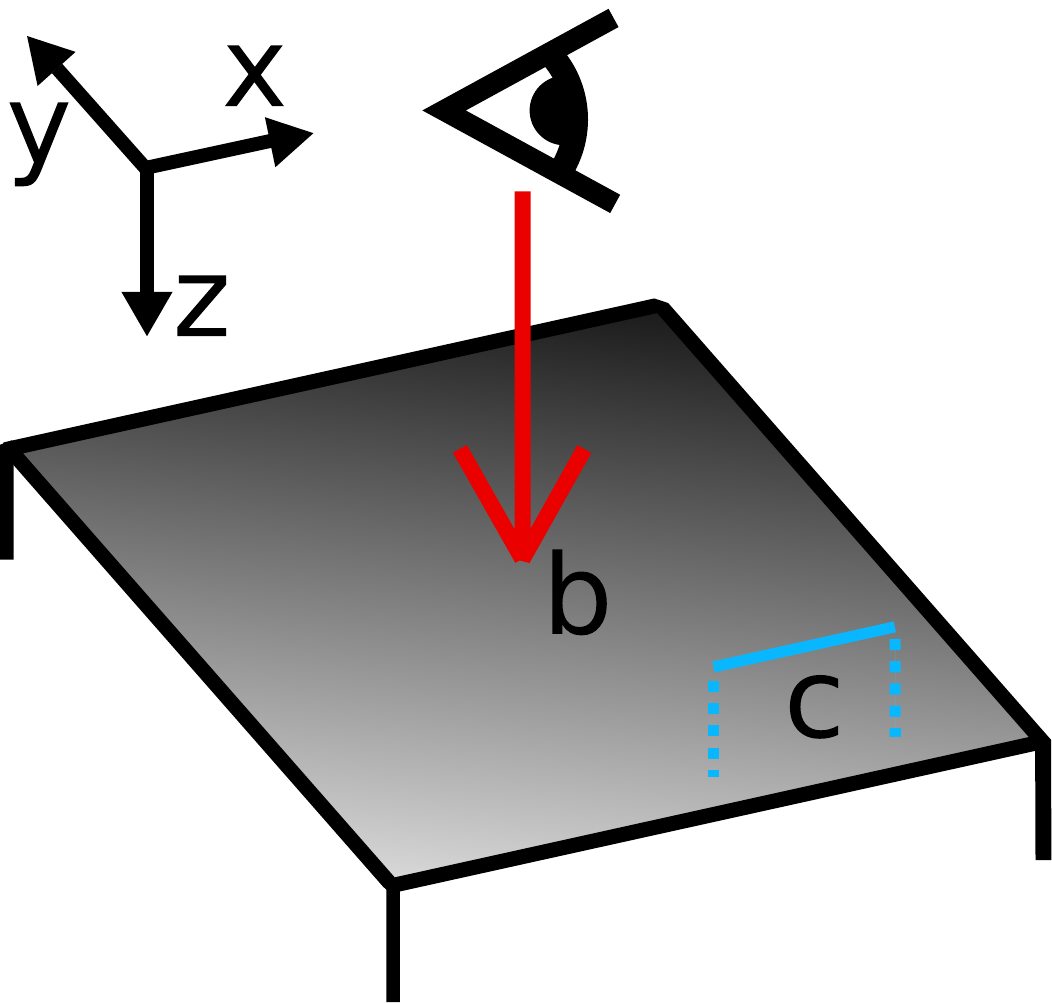}
        \caption{Model texture}
    \end{subfigure}
    \begin{subfigure}[t]{0.26\columnwidth}
        \centering        
        \includegraphics[width=\columnwidth]{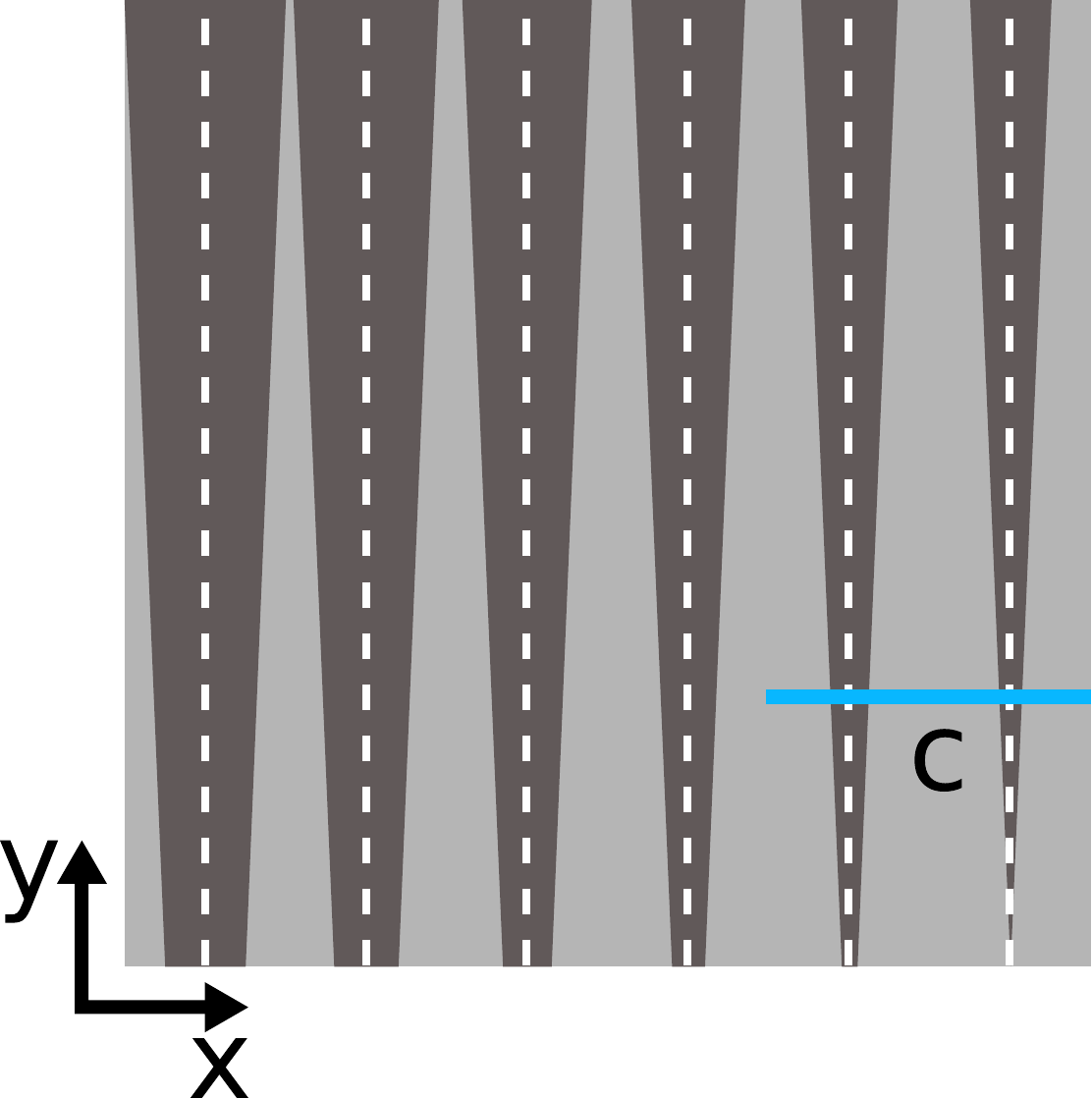}
        \caption{Top view}
        \label{fig:linear_halftoning_top}
    \end{subfigure}
    \begin{subfigure}[t]{0.45\columnwidth}
        \centering        
        \includegraphics[width=\columnwidth]{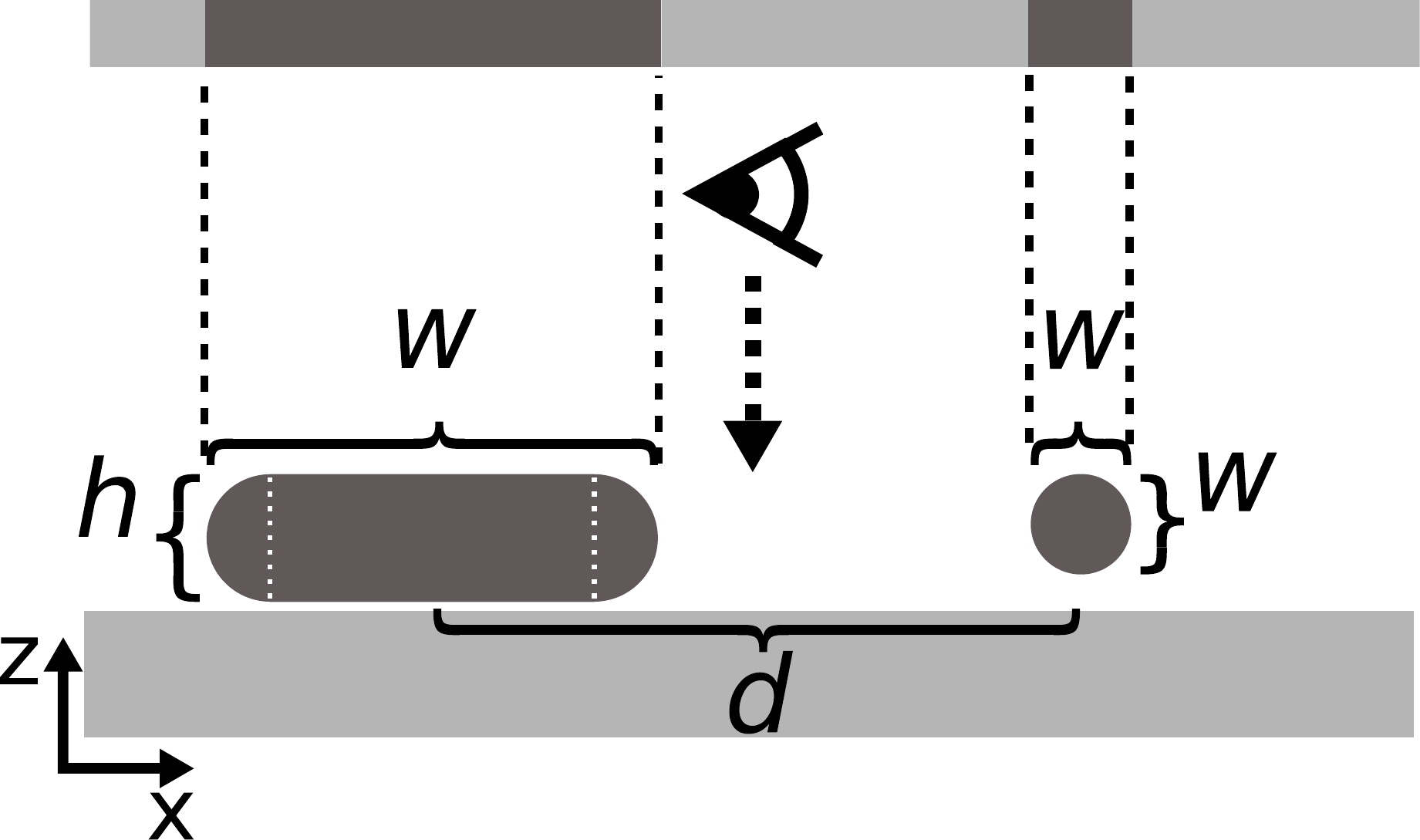}
        \caption{Cross-section}
        \label{fig:line_cross_section}
    \end{subfigure}
    \caption{
        Horizontal hatching of a linear gradient.
        The 'c' marks where the cross section of figure~\subref{fig:line_cross_section} is located.
        The dashed lines in \subref{fig:linear_halftoning_top} show the travel paths of the center of the nozzle while printing the black lines.
        \subref{fig:line_cross_section} shows the model used for achieving lines of a given width.
    }
    \label{fig:horizontal_hatching}
\end{figure}

The presented hatching technique makes use of the stair-stepping effect and sagging, but these don't occur on horizontal surfaces.
Therefore, horizontal top and bottom surfaces of 3D meshes are instead hatched by modulating the widths of the lines used to print the top/bottom skin.
The most commonly used pattern to fill in these areas is a regular grid of equidistant straight lines. 
Grayscale gradients are achieved by modulating the width of these lines.
When printing a black layer the line widths are modulated so that parts of the adjacent white layer become visible and vice versa.
A visual representation of this is presented in figure~\ref{fig:linear_halftoning_top}.
The resulting halftoning images resemble the image in figure~\ref{fig:engraving_style}.

The texture image is sampled at regular intervals along the skin lines and the width $w$ of each line segment between two consecutive sample points is determined by the line distance $d$ and the average texture luminance $L$, which is in the range $(0,1)$: $w = Ld$.

Varying the width of lines is achieved by increasing and decreasing the amount of deposited material.
For Bowden style FDM printers it takes relatively long to change the amount of material departing the nozzle per second - a.k.a. the \emph{flow}.
Therefore we propose keeping the flow constant and varying the movement speed $v$ of the print head rather than varying the flow at a constant movement speed: 
$v = c / A$, where $A$ is the area of a cross section of the printed line and $c$ is an empirically determined constant flow.
We model the cross section as a rectangle with semicircular sides: $A = \pi (0.5h)^2 + h(w-h)$, where $h$ is the layer thickness.
See figure~\ref{fig:line_cross_section}.
For lines narrower than the layer thickness we model the cross section as a circle: if $w < h$ then $A = \pi(0.5w)^2$.


\subsection{Implementation}
This section describes the implementation of the variable offsets required for vertical and diagonal hatching.

\subsubsection{Texture Mapping} \label{section:texture_mapping}
Before the variable offsets are applied, the model textures are mapped onto the outlines of each layer.
This is done in the phase where the surface mesh is sliced into outlines; for each face which produces a line segment of an outline, the corresponding texture line segment is recorded.
Each location on the polygons of the outlines of a layer can then be mapped to a UV coordinate of the texture image.
The grayscale value at the point in the image is then used to calculate by which distance the outline is displaced at the location.





\begin{figure}
    \centering
    \begin{subfigure}[b]{0.53\columnwidth}
        \centering
         \includegraphics[width=\columnwidth]{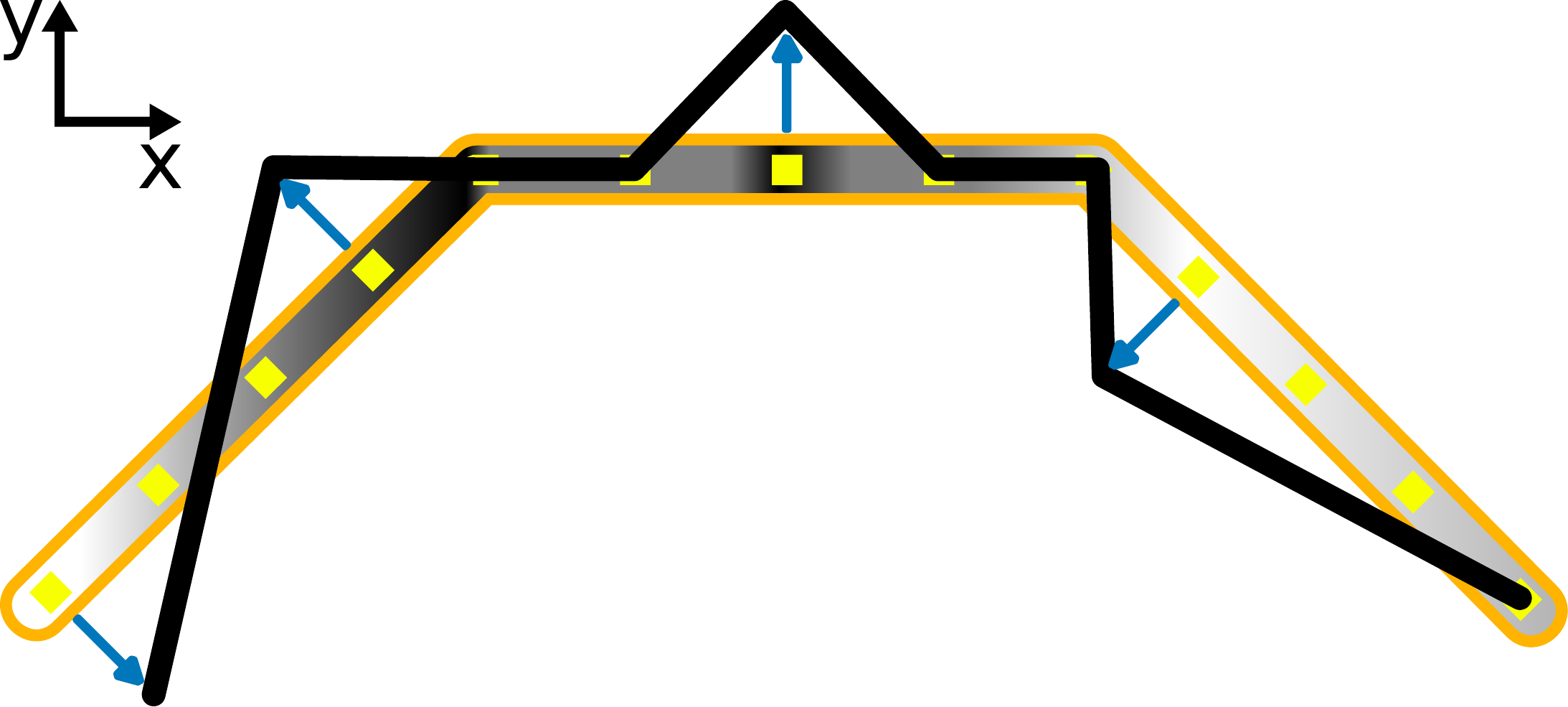}
        \caption{Variable offset for vertical hatching}
        \label{fig:variable_offset_general}
    \end{subfigure}
    \begin{subfigure}[b]{0.22\columnwidth}
        \centering
         \includegraphics[width=\columnwidth]{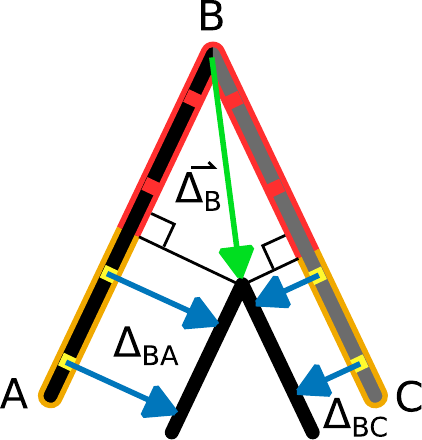}
        \caption{Shortcut corner}
        \label{fig:variable_offset_corner_in}
    \end{subfigure}
    \begin{subfigure}[b]{0.2\columnwidth}
        \centering
         \includegraphics[width=.75\columnwidth]{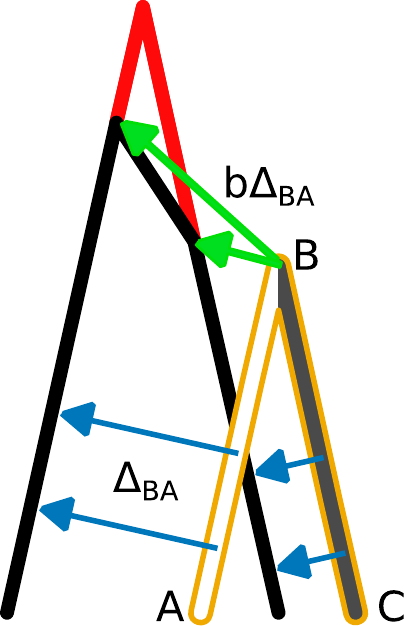}
        \caption{Beveled corner}
        \label{fig:variable_offset_corner_out}
    \end{subfigure}
    \caption{
        Visual explanation of variable offset applied to parts of a polygon.
        The original polygon is shown in orange,
        the resulting offset polygon in black,
        the sampling points in yellow
        and the offsets in blue.
        The grayscale values determine the amplitude of the offset.
        The green offsets show how variable offsetting of corners should be handled.
        The red items are omitted from the end result.
    }
    \label{fig:variable_offset}
\end{figure}

\begin{figure*}[t!]
    \centering
    \begin{subfigure}{0.16\textwidth} \centering \includegraphics[height=\textwidth]{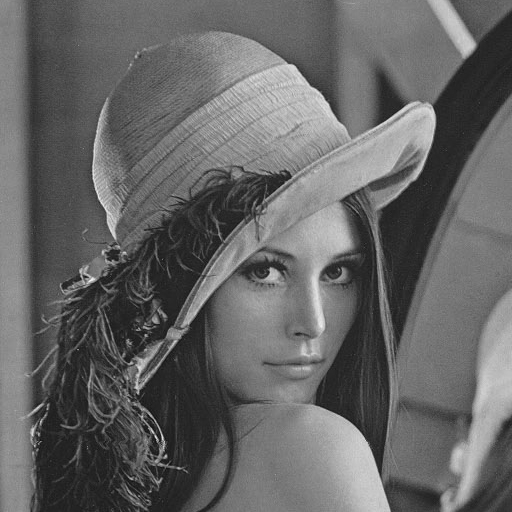} \caption{Input} \label{fig:lena_input} \end{subfigure}
    \begin{subfigure}{0.16\textwidth} \centering \includegraphics[height=\textwidth]{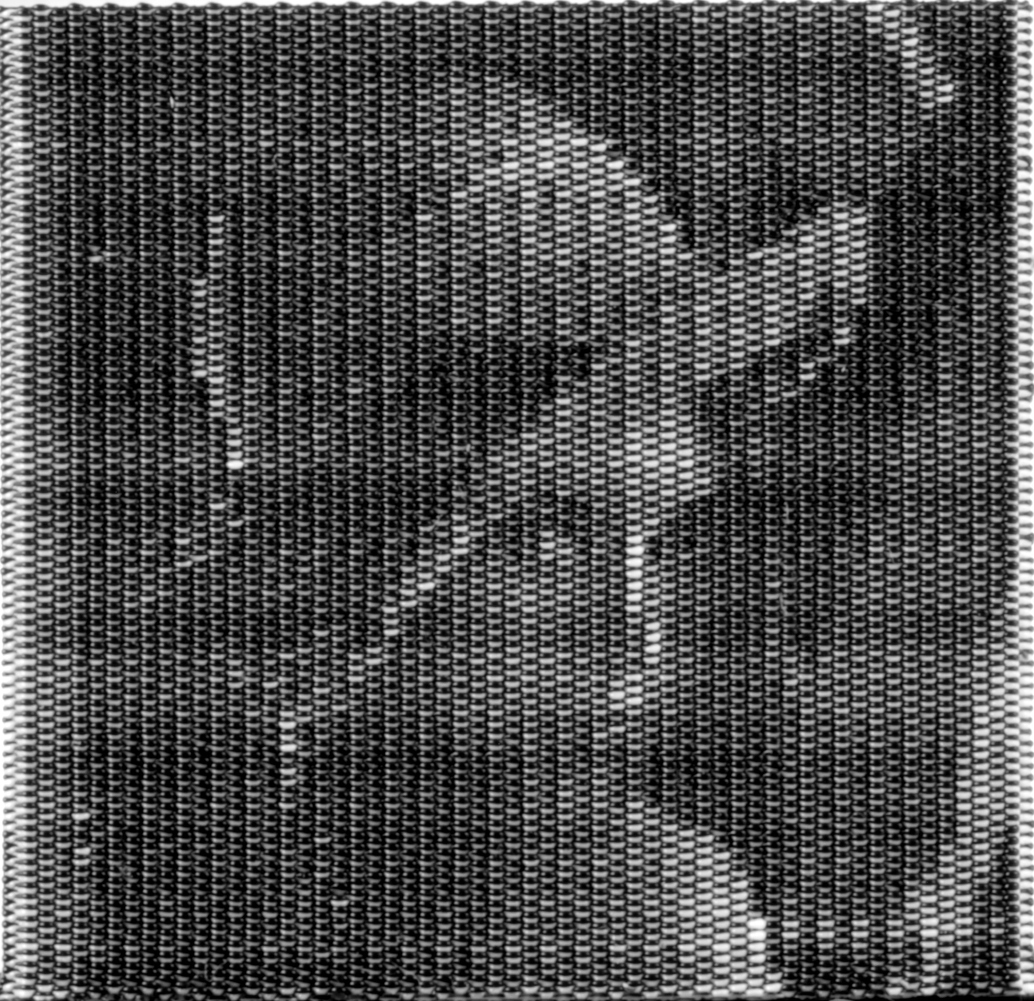} \caption{\citeauthor{reiner2014dual}\cite{reiner2014dual}} \label{fig:lena_reiner} \end{subfigure}
    \begin{subfigure}{0.16\textwidth} \centering \includegraphics[height=\textwidth]{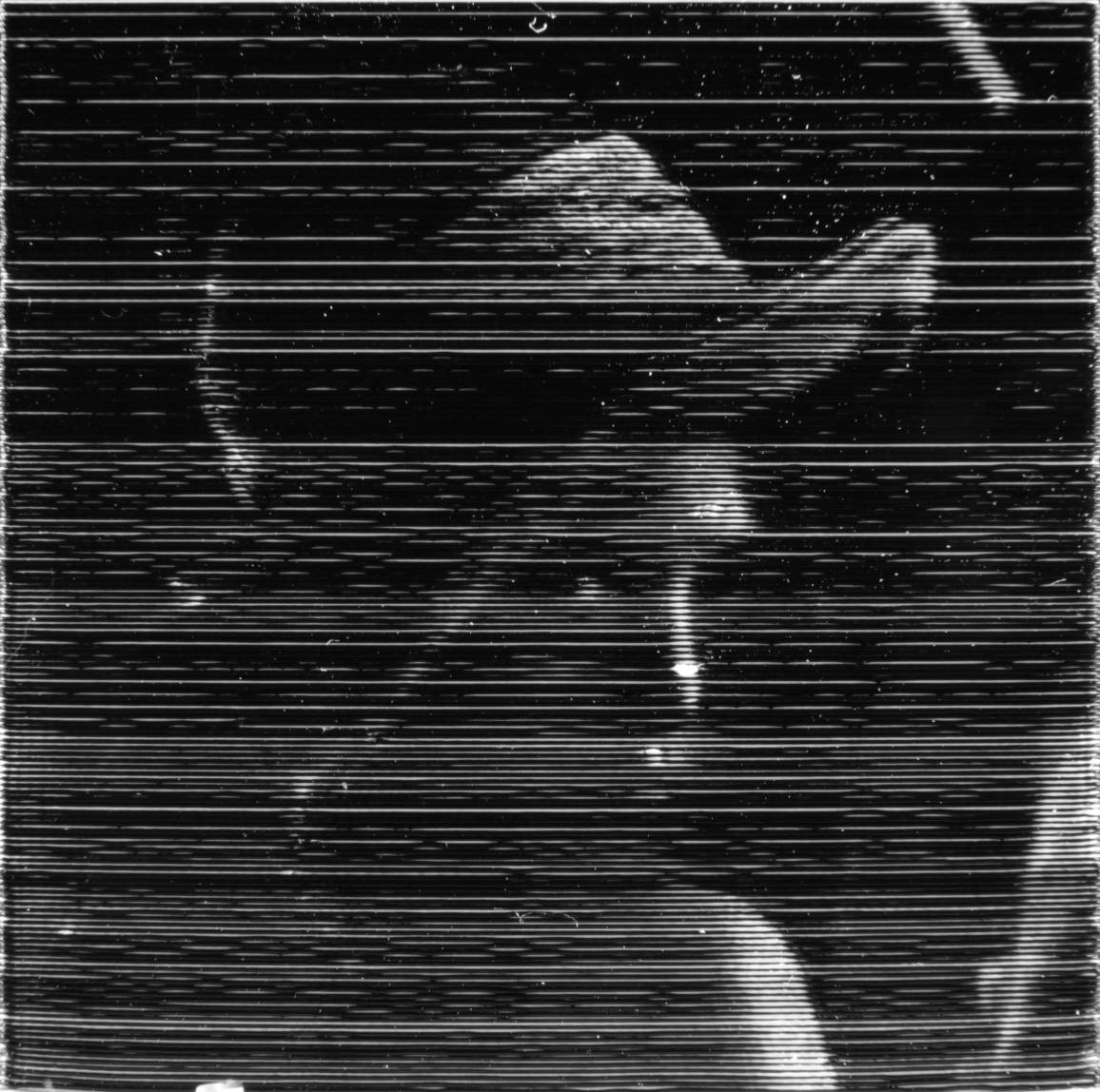} \caption{\SI{0}{\degree}}   \label{fig:lena_00} \end{subfigure}
    \begin{subfigure}{0.16\textwidth} \centering \includegraphics[height=\textwidth]{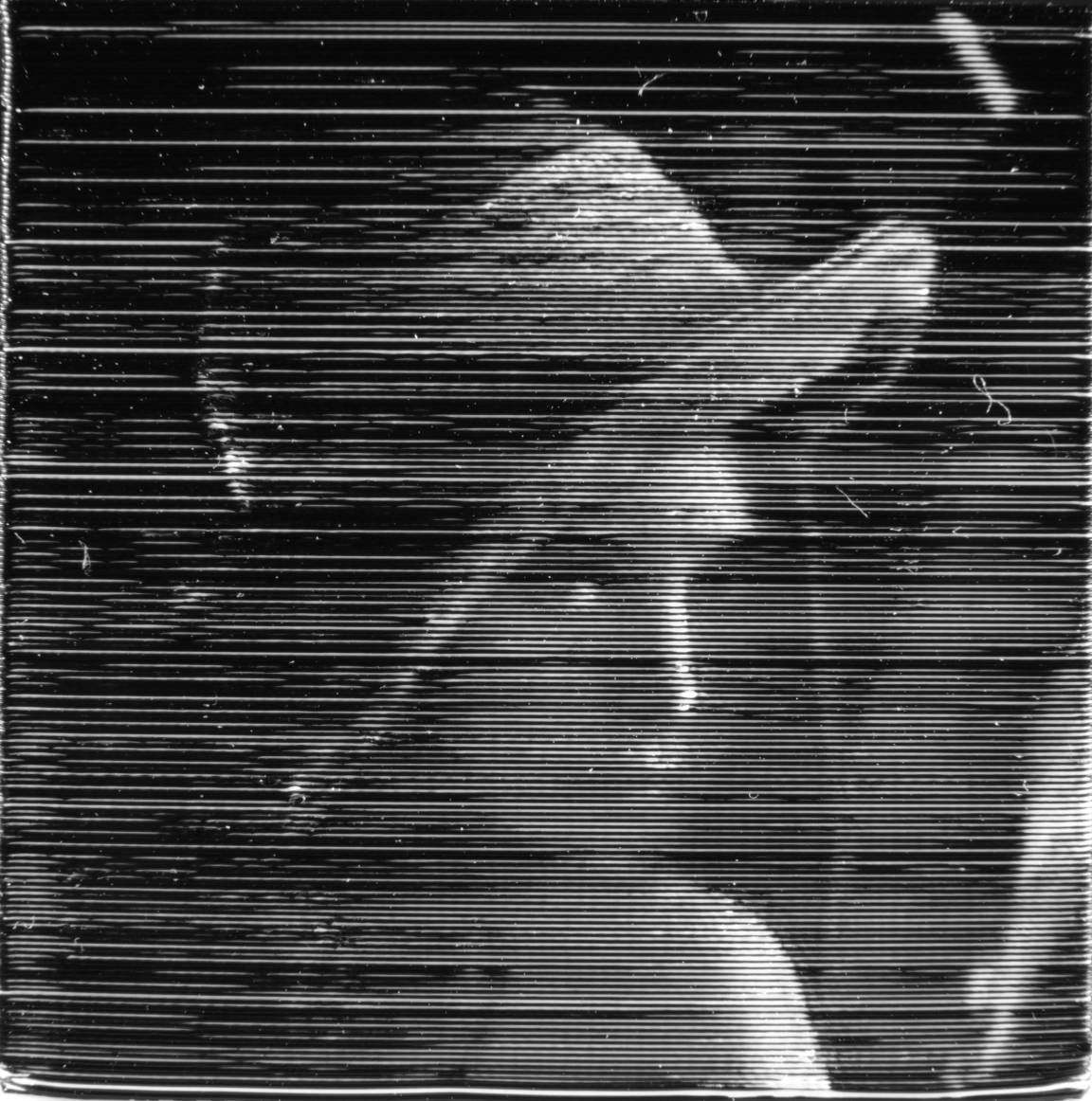} \caption{\SI{10}{\degree}}  \label{fig:lena_10} \end{subfigure}
    \begin{subfigure}{0.16\textwidth} \centering \includegraphics[height=\textwidth]{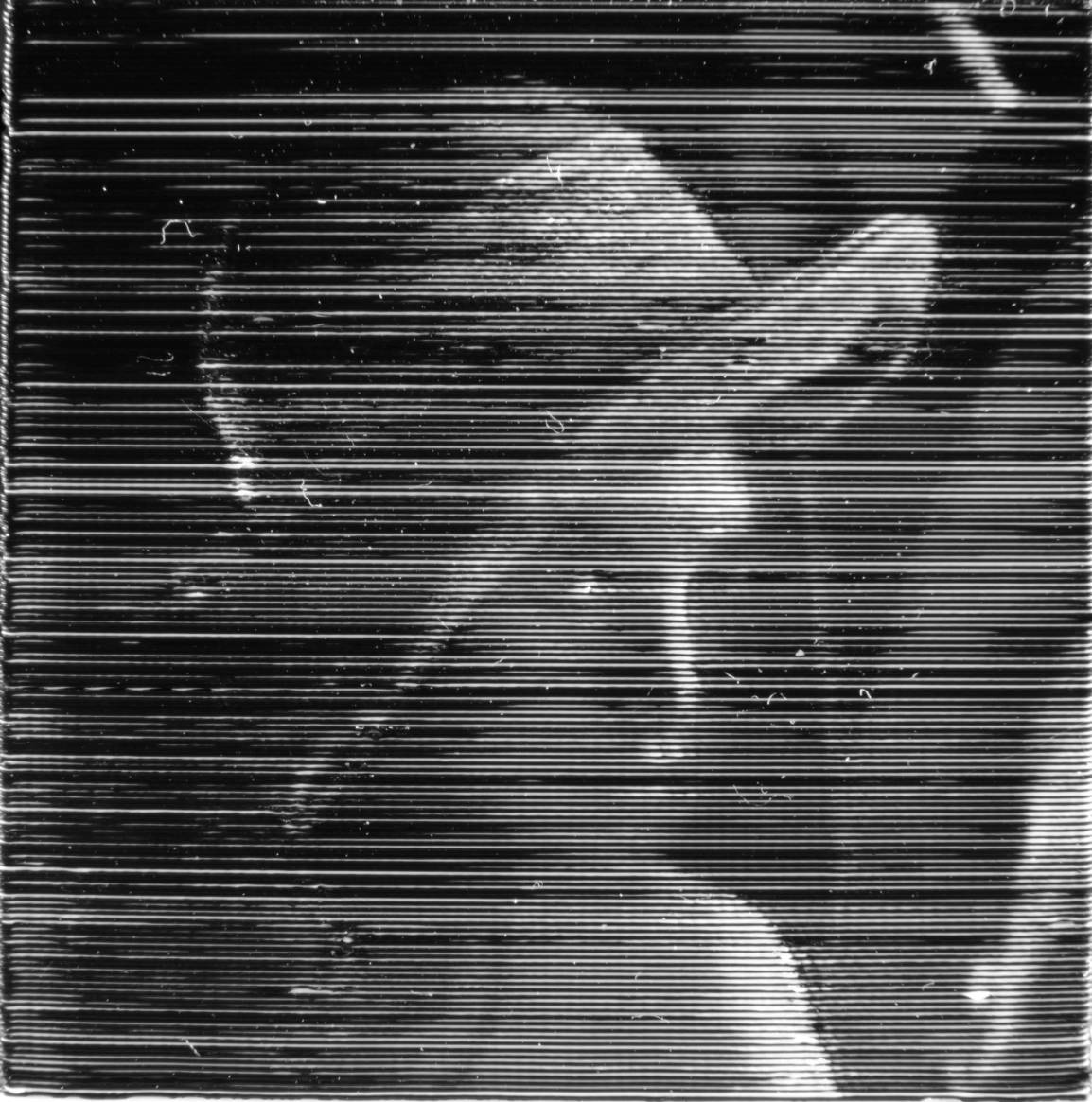} \caption{\SI{20}{\degree}}  \label{fig:lena_20} \end{subfigure}
    \begin{subfigure}{0.16\textwidth} \centering \includegraphics[height=\textwidth]{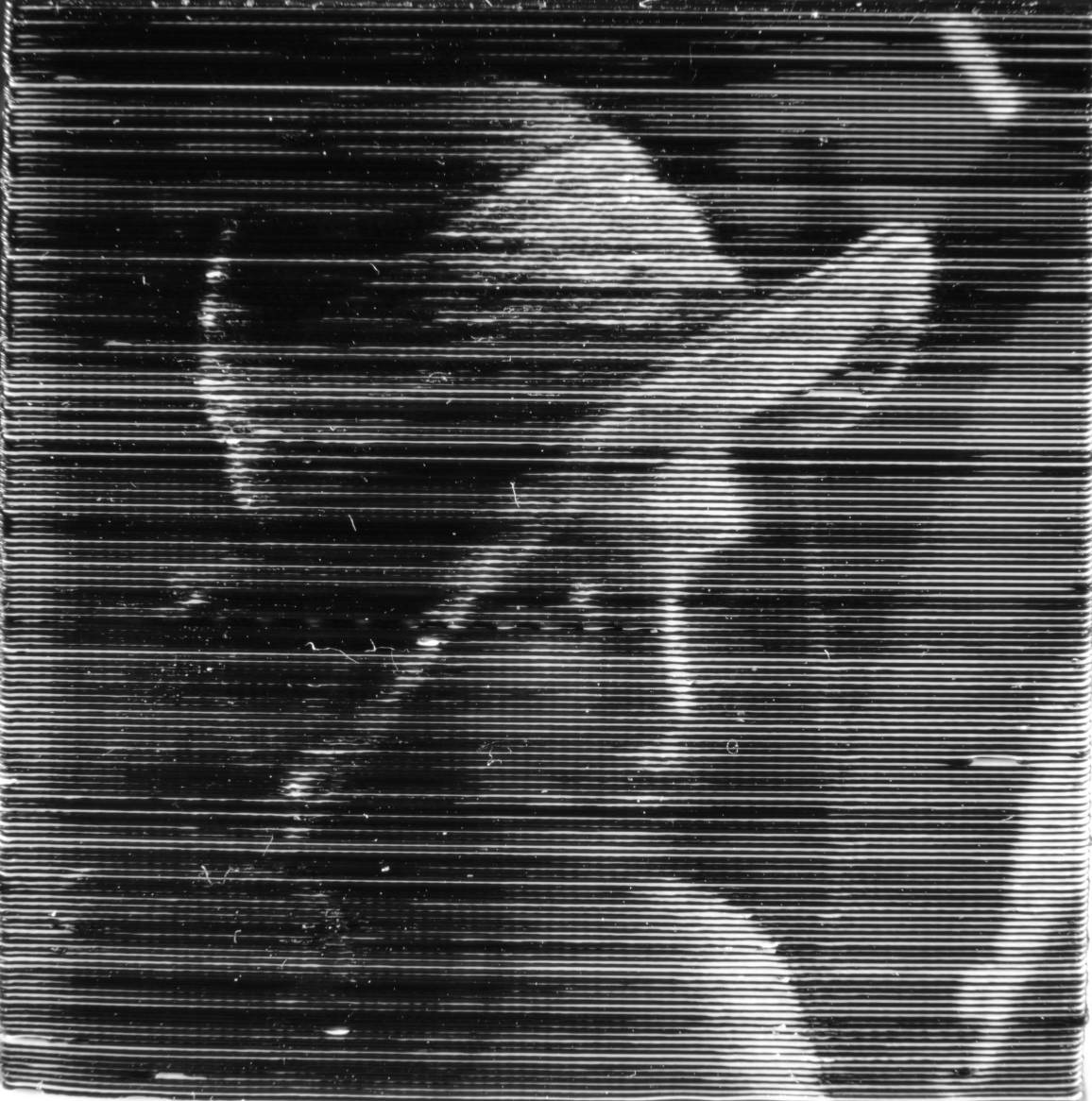} \caption{\SI{30}{\degree}}  \label{fig:lena_30} \end{subfigure}
    \begin{subfigure}{0.16\textwidth} \centering \includegraphics[height=\textwidth]{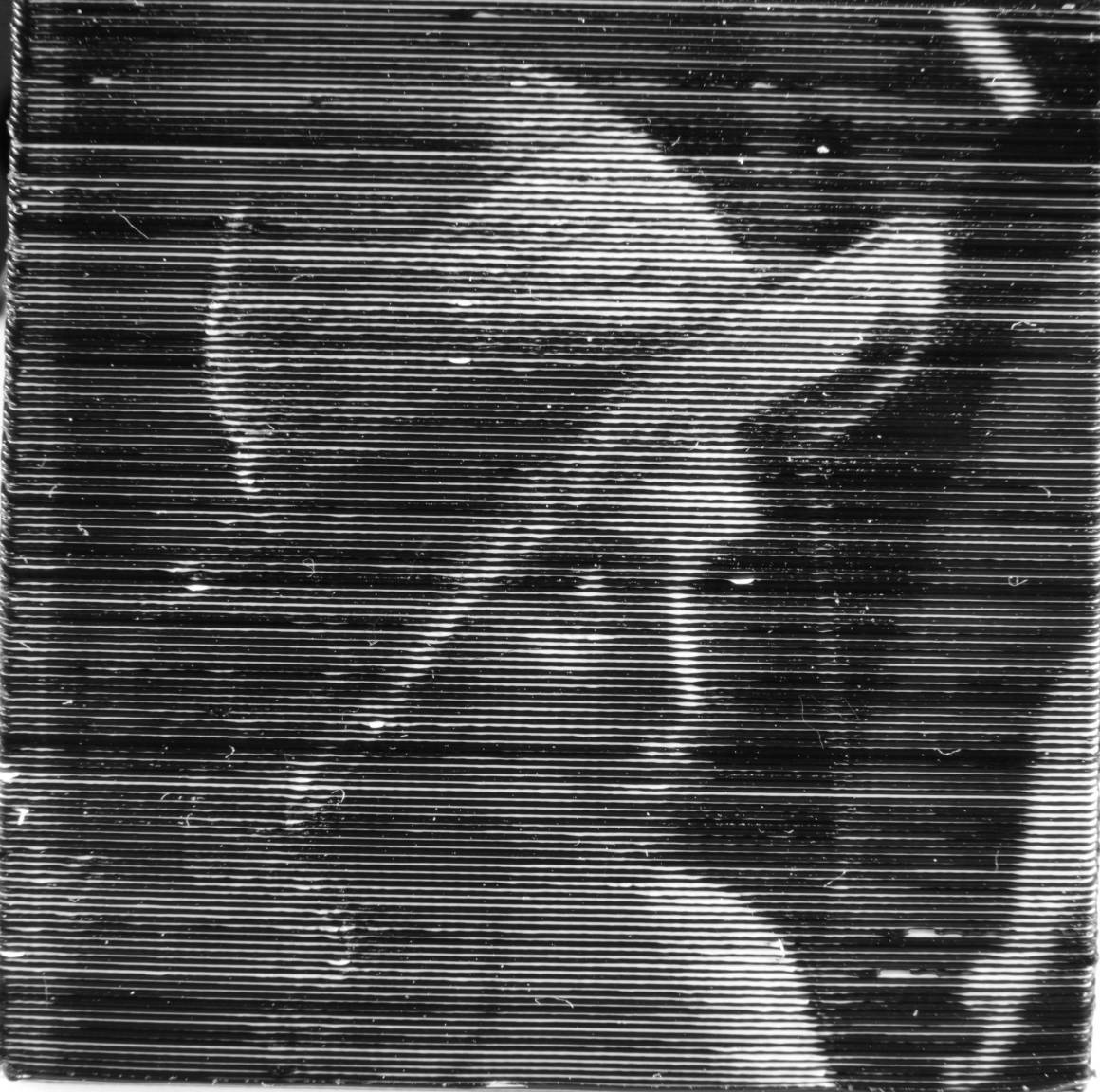} \caption{\SI{40}{\degree}}  \label{fig:lena_40} \end{subfigure}
    \begin{subfigure}{0.16\textwidth} \centering \includegraphics[height=\textwidth]{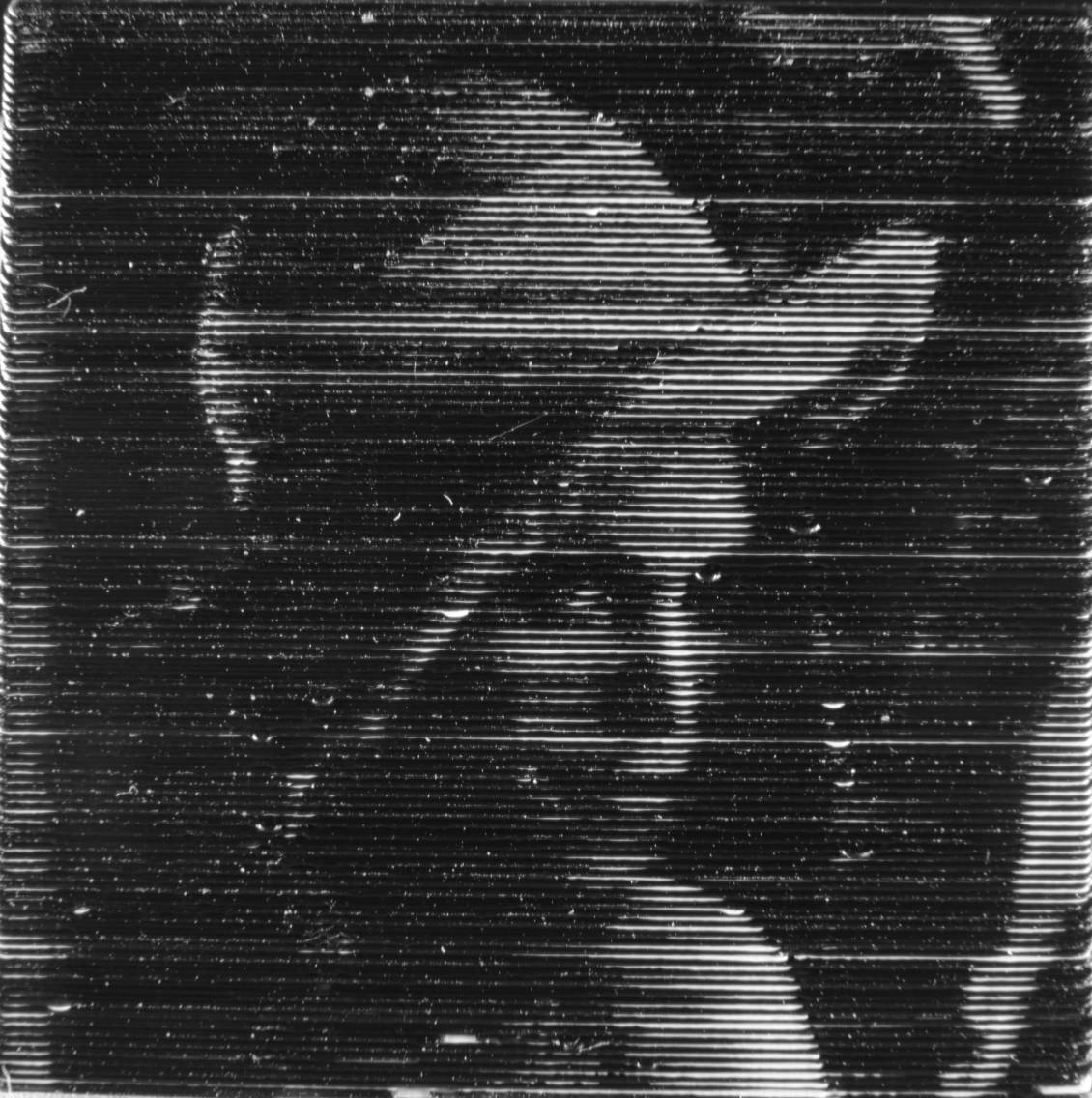} \caption{\SI{50}{\degree}}  \label{fig:lena_50} \end{subfigure}
    \begin{subfigure}{0.16\textwidth} \centering \includegraphics[height=\textwidth]{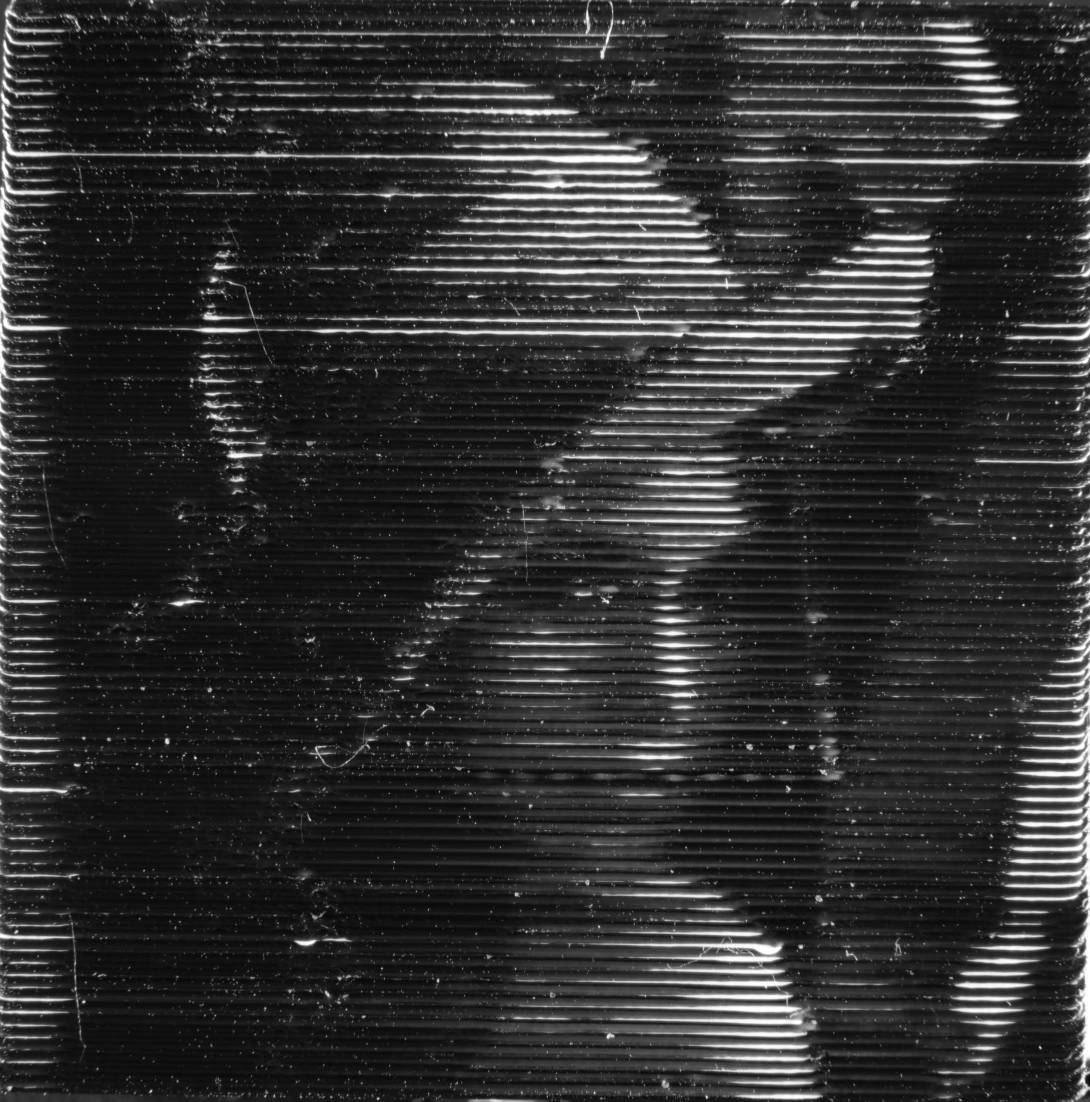} \caption{\SI{60}{\degree}}  \label{fig:lena_60} \end{subfigure}
    \begin{subfigure}{0.16\textwidth} \centering \includegraphics[height=\textwidth]{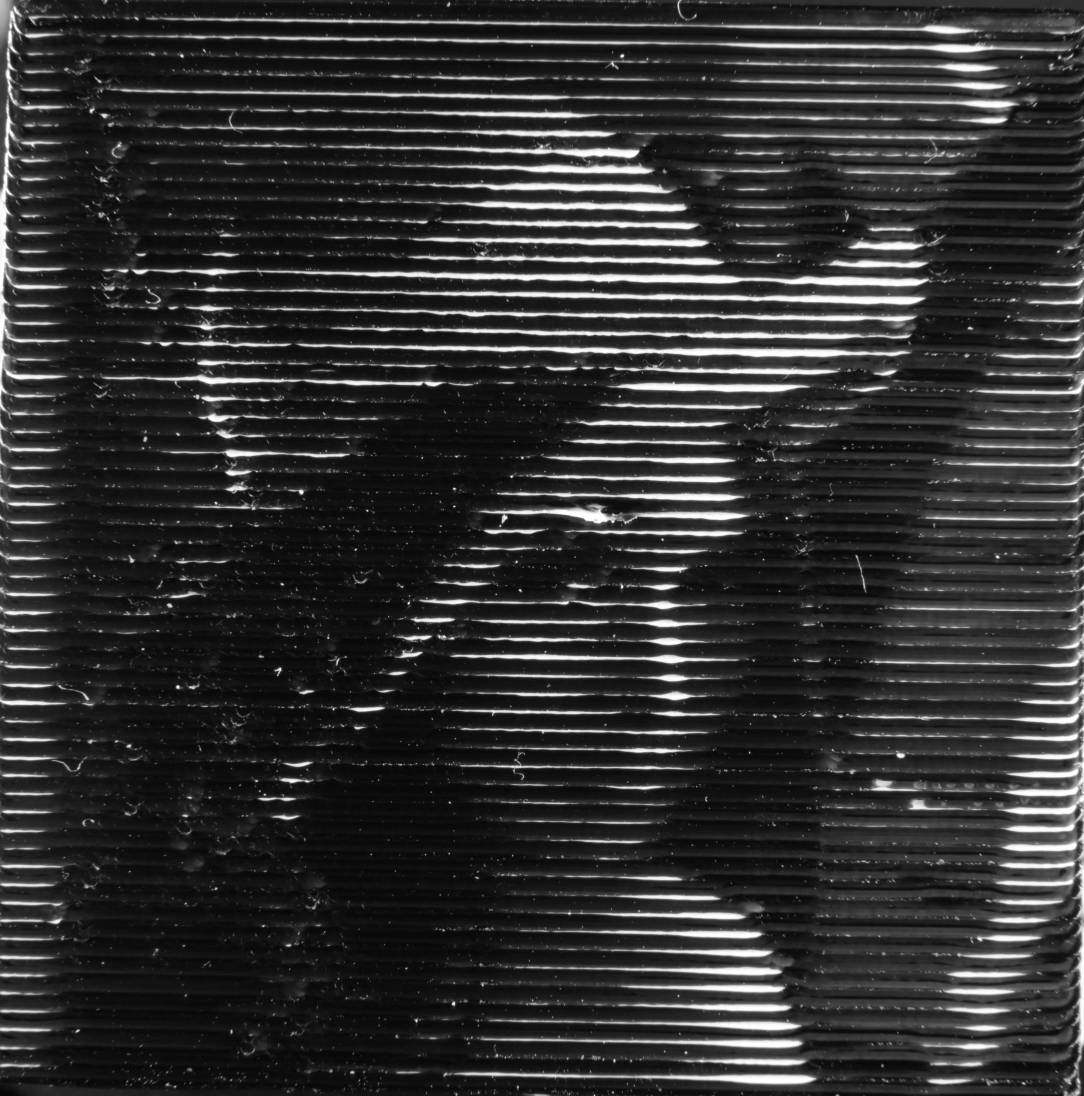} \caption{\SI{70}{\degree}}  \label{fig:lena_70} \end{subfigure}
    \begin{subfigure}{0.16\textwidth} \centering \includegraphics[height=\textwidth]{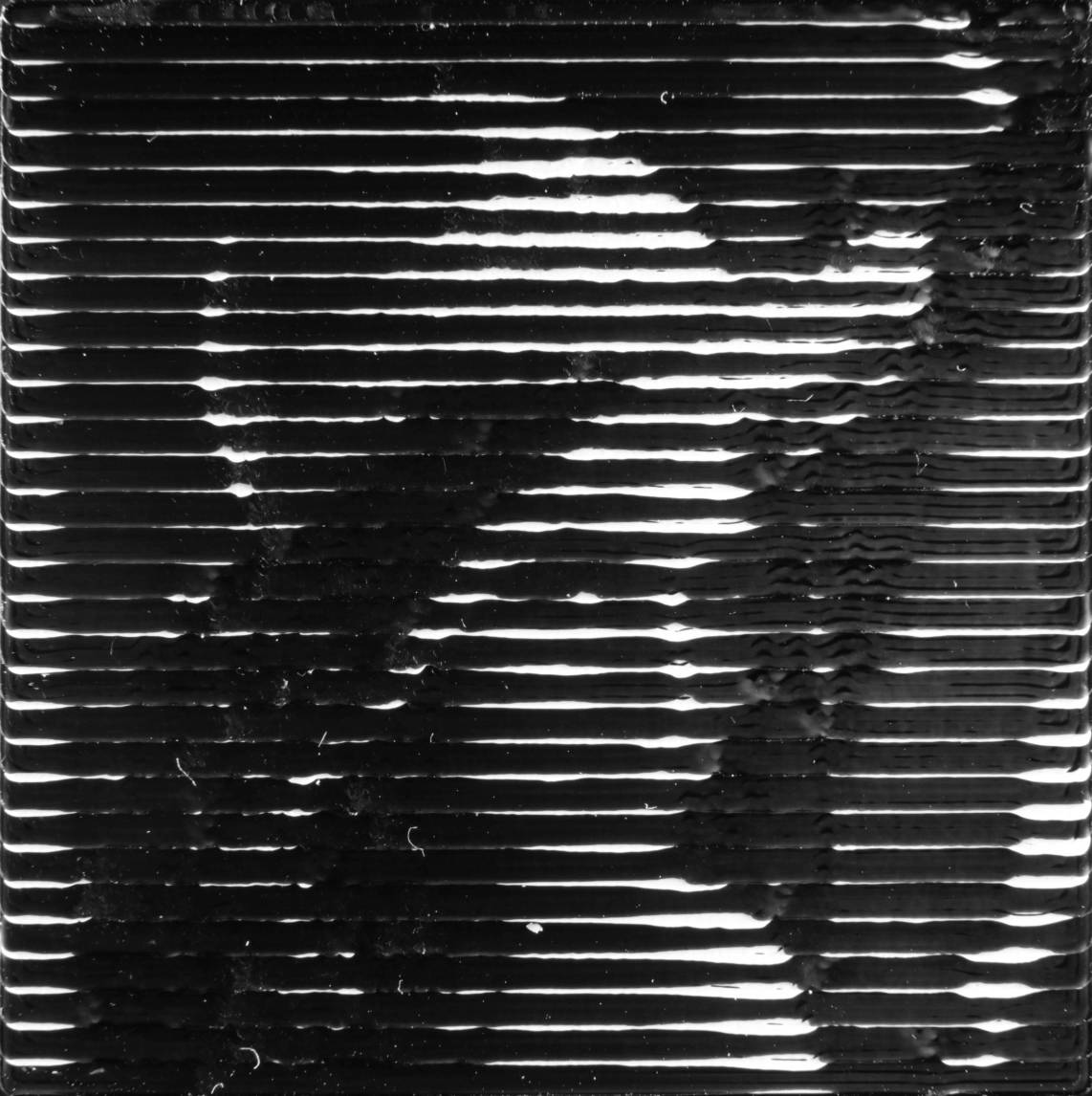} \caption{\SI{80}{\degree}}  \label{fig:lena_80} \end{subfigure}
    \begin{subfigure}{0.16\textwidth} \centering \includegraphics[height=\textwidth]{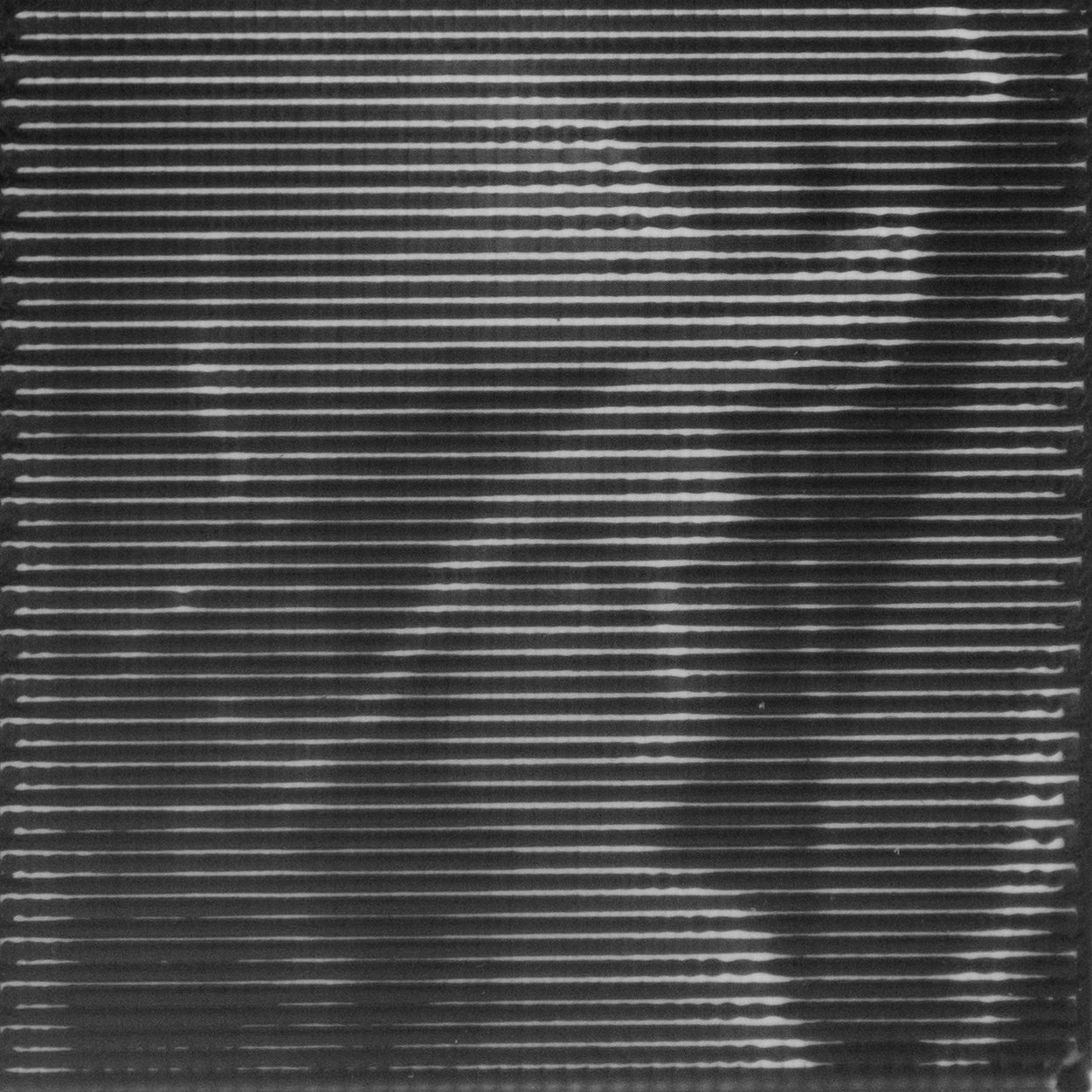} \caption{Horizontal} \label{fig:lena_horizontal} \end{subfigure} 
    \caption{A $\SI{35}{}\times \SI{35}{\milli\meter}$ printed surface for various surface angles $n$ optimized for a perpendicular viewing angle.
Figure~\subref{fig:lena_reiner} shows a reproduction of the method by \citeauthor{reiner2014dual} on a vertical surface without tone calibration.
Figure~\subref{fig:lena_horizontal} shows the horizontal hatching technique.
Image courtesy of \textcopyright Playboy Magazine  1972.
    }
    \label{fig:lena}
\end{figure*}

\subsubsection{Variable Offset}
The displacement given by the equations above is used to generate a variable offset polygon from the existing outlines of a layer.
Both vertices and intermediate points on line segments are displaced in the outward direction perpendicular to the polygon.
Line segments are offset by displacing points sampled at a regular interval, as can be seen in figure~\ref{fig:variable_offset_general}.



Given that vertices in the outline generally belong to an edge of the 3D model connecting \emph{two} faces, which can have different texture image locations at that vertex, applying a variable offset should take two displacement values into account.
The offset $\vec{\Delta_B}$ applied to the vertex at the corner between two line segments ${BA}$ and ${BC}$ is given by
\begin{equation}
\vec{\Delta_B} = \frac{ \Delta_{BA} \abs{\vec{BA}} \vec{BC} + \Delta_{BC} \abs{\vec{BC}} \vec{BA}}
     { \det \left[ \vec{BA}^T \vec{BC}^T \right] }
\end{equation}
, where $\Delta_{BA}$ is the offset at $B$ which follows from the texture coordinates at $B$ on the mesh face which $AB$ is located on and likewise for $\Delta_{BC}$.


For inward offsets like the one shown in figure~\ref{fig:variable_offset_corner_in}, the offset corner bypasses part of the connected line segments.
Projecting the offset $\vec{\Delta_B}$ onto the line segments $BA$ and $BC$ gives the distance by which we disregard the sampling points along those line segments.
The disregarded sampling points on the edges are not displaced; they are omitted from the variably offset polygon.

When using the above formula, sharp corners which take an outward displacement into account, could result in corners which are displaced by a distance greatly exceeding either displacement value.
The resulting corner is therefore beveled when the displacement $\vec{\Delta_{B}}$ is larger than the bevel distance of both line segments.
The bevel distance of a line segment $\vec{BA}$ is given by $b\Delta_{BA}$, where $b>1.0$ is a constant ratio which determines how much of the corner is beveled off.
In figure~\ref{fig:variable_offset_corner_out} a bevel ratio $b=1.1$ was used.

Where line segments in the original outlines are close to each other and the offsets applied are larger than the distance between the line segments, the variably offset polygon would contain self-intersections.
The self-intersecting parts of the polygon should be removed by applying a polygon clipping operation which uses a filling rule based on a positive winding number.
For further reading, see \citet{vatti1992generic}.



\section{Results}
Experiments were performed on an Ultimaker 3 machine, using black and white polylactic acid (PLA) - resp.\  Ultimaker 9014 and Ultimaker 9016.
The print cores used in this setup have a nozzle size of \SI{0.4}{\milli\meter} and a line width setting of \SI{0.35}{\milli\meter} was used.
A default layer thickness of \SI{0.1}{\milli\meter} was used.
Different head movement speeds were used throughout the printing process; most notably, the outer walls were printed with a speed of \SI{15}{\milli\meter\per\second}.
In order to follow the outline as accurately as possible, the outer wall was printed before the inner walls.
We used a sampling distance of \SI{0.1}{\milli\meter}.
Furthermore we applied a static offset of $\SI{0.1}{\milli\meter}$ to all outlines in order to prevent self-intersections due to negative offsets on either side of a thin outline polygon.

In an initial testing phase of sagging behavior multiple overhang distances $o$ were tested at \SI{0.01}{\milli\meter} intervals.
After these experiments it seemed that applying an offset of \SI{0.2}{\milli\meter} yields satisfactory results: \SI{0.2}{\milli\meter} was the minimal tested offset at which black pixels in the texture appeared fully black when viewed from the side and vice versa for white.
The sagging ratio was therefore estimated at $w/h = 2.0$, which was then used along with the geometrical sagging model to determine the offsets from texture color information.

The standard Lena test image (figure~\ref{fig:lena_input}) has been printed on surfaces of different slope using our hatching model optimized for perpendicular viewing.
The results can be seen in figure~\ref{fig:lena}.

Tests on horizontal hatching culminated in a result which can be seen in figure~\ref{fig:lena_horizontal}, which shows the top of a test print containing a single wall and skin lines.
The horizontal hatching technique was performed on the top black layer with a fully dense white layer below.
The texture was sampled at \SI{0.4}{\milli\meter} intervals.
A line distance of \SI{0.7}{\milli\meter} was used, a reference speed of \SI{25}{\milli\meter\per\second} and a reference line width of \SI{0.35}{\milli\meter} to produce the constant flow of \SI{0.875}{\milli\meter\cubed\per\second}.

Figure~\ref{fig:teaser} shows hatched objects with geometric and organic shapes and with low and high frequency texture detail.
The time it takes to print these objects using the 3D hatching techniques was up to $15\%$ longer when compared to printing the same models with a single extruder.
See table~\ref{tab:timing}.
This was unforeseen, provided that switching extruder on the Ultimaker 3 typically takes up less than $1\%$ of the print time.
We postulate that the difference in printing time is caused by the irregularity introduced by the variable offset.
This irregularity causes the print head movement speed to be limited by its acceleration settings.

Because our hatching algorithms change the outlines by offsetting points sampled each \SI{0.1}{\milli\meter}, the number of points in the outlines increases drastically.
This further impacts a lot of the other process planning algorithms down the line.
We can see in table~\ref{tab:timing} that applying  hatching to the rhino figurine only added \SI{20}{\percent} extra process planning time, while that caused an added slice time of \SI{70}{\percent} down the line.

\begin{table}
\caption{Comparison of slice times of the total process planning, processing time spent on performing the variable offset alone and print times between normal FDM printing and our hatching technique on the prints presented in figure~\ref{fig:teaser}.}
\label{tab:timing}
\begin{tabular}{l c c c  c c }
  & \multicolumn{2}{ | c }{normal} & \multicolumn{3}{ | c  }{hatching}  \\
model   & \multicolumn{1}{ | c }{slice}  & print&  \multicolumn{1}{ | c }{offset} & slice  & print  \\
\hline 
Figurine    & \SI{158}{\second} & \SI{32}{\hour}\SI{37}{\minute}     & \SI{32}{\second} & \SI{300}{\second}      & \SI{36}{\hour}\SI{8}{\minute}     \\ 
Portrait     & \SI{69}{\second} & \SI{23}{\hour}\SI{30}{\minute}      & \SI{9}{\second}  & \SI{88}{\second}       & \SI{24}{\hour}\SI{20}{\minute}     \\ 
Can  & \SI{34}{\second} & \SI{16}{\hour}\SI{28}{\minute}      & \SI{6}{\second}  & \SI{47}{\second}       & \SI{17}{\hour}\SI{9}{\minute}     \\  
Rod      & \SI{16}{\second} & \SI{6}{\hour}\SI{35}{\minute}       & \SI{4}{\second}  & \SI{21}{\second}       & \SI{7}{\hour}\SI{29}{\minute}     
\end{tabular}
\end{table}


\section{Discussion}
\subsection{Texture resolution}
Figure~\ref{fig:lena} shows the texture resolution obtained with the 3D hatching technique, compared to the input image (~\ref{fig:lena_input}) and the technique of \citet{reiner2014dual} (~\ref{fig:lena_reiner}).
Figure~\ref{fig:lena_00} show a vertical print hatched for horizontal viewing.
The vertical resolution depends on the layer thickness, which was \SI{0.1}{\milli\meter} for all prints.
The results obtained using vertical halftoning have a higher horizontal resolution than results obtained using the method by \citeauthor{reiner2014dual}: figure~\ref{fig:lena_reiner}.
Because vertical hatching is not limited by the wave length, the resolution of the resulting print is not limited by the line width, which is related to the physical size of the hole in the nozzle.

Figure~\ref{fig:lena_reiner} was created using the technique from \citet{reiner2014dual}.
A wave length of \SI{1.2}{\milli\meter} was used; a minimum amplitude of \SI{0.375}{\milli\meter} and a maximum amplitude of \SI{0.75}{\milli\meter}.
Because the object is a simple cube for which each slice has the same square cross-section, projection threshold and relaxation don't play a role.
No tone calibration has been performed; the amplitude depends linearly on the luminance of the texture input.
The implementation is included in \citet{cura}.

Figures~\ref{fig:lena_00}~to~\subref{fig:lena_80} show result obtained using hatching optimized for perpendicular viewing on surfaces with various surface slopes $n$. 
The vertical resolution decreases with higher surface angles $n$, because the stair-step width gets larger.
However, the grayscale tones remain roughly the same across the different surface slopes --- notwithstanding the lack of tone calibration and the presence of visual artifacts.

Horizontal hatching shows a relatively low resolution.
This is because the resolution is determined predominantly by the nozzle size, which was high in comparison to the layer thickness in our tests.
Because the black lines sit on top of a white layer, the perceived tone tends toward black when viewing the surface from a lower altitude. 
Close examination of figure~\ref{fig:lena_horizontal} reveals that at places where the luminance value of the texture is high, black lines appear to be printed thinner, rather than narrower.
It seems that these thinner lines only partially block light from the previous layer, resulting in a grayscale value close to the texture luminance even though their width is larger than intended.

It should be noted that all of the results in figure~\ref{fig:lena} seem darker than the input image.
This might be explained by several factors which affect the perceived luminance.
Several such factors are mentioned in the remainder of this section.

\subsection{Dimensional accuracy and slicing}
Because the presented halftoning technique works by altering the geometry of a layer it affects the dimensional accuracy of the print.
The size of the offset is always below $0.5 d+w$, i.e. it's smaller than the stair step width aside for the extra offset required for sagging.
This means that if it weren't for sagging, the dimensional accuracy of the resulting object would be at least as good as the dimensional accuracy of a print job performed with twice the layer thickness.
The maximum distance from the surface mesh to the surface of the printed product incurred by the stair-stepping effect is half the layer thickness.
For our tests with a layer thickness of $h=\SI{0.1}{\milli\meter}$ the range in which our technique employs the sagging effect is $0.5 w=\SI{0.1}{\milli\meter}$.
Since for perpendicular viewing the sagging is used when the stair step are narrow and less so when the stair steps are wide, 
the maximum dimensional error incurred is \SI{0.1}{\milli\meter}.
For near horizontal surfaces this inaccuracy incurred drops to \SI{0.05}{\milli\meter} on top of the existing inaccuracy due to the stair-stepping effect.

The proposed hatching technique incurs minimal changes to the patterns by which conventional slicing software applications generate semi-continuous lines to build up a 3D printed object.
The layer thickness remains unaltered and the layers are still built up by the walls, skin and infill.
Keeping the elementary properties of FDM unaltered means that the structural properties of 3D prints are minimally influenced.

\subsection{Visual artifacts}
Variations in print parameters and inaccuracies in the printing system lead to various (visual) artifacts. 
These artifacts are particularly pronounced in hatching, due to the high contrast between consecutive layers. 
These artifacts cannot be well described with existing quality metrics for FDM, like (global) geometric accuracy and/or surface roughness (i.e. \cite{Armillotta2017a,Kim2018}), both which do not capture the effect of the geometric variation on the perceived gray scale.
Even though the sagging of vertical surfaces is employed within the range of \SI{0.2}{\milli\meter}, we found the human eye can easily detect these inaccuracies, especially in regions with a uniform gray scale. 

\begin{figure}
    \centering
    \begin{subfigure}[b]{0.19\columnwidth}
        \centering
         \includegraphics[width=\textwidth]{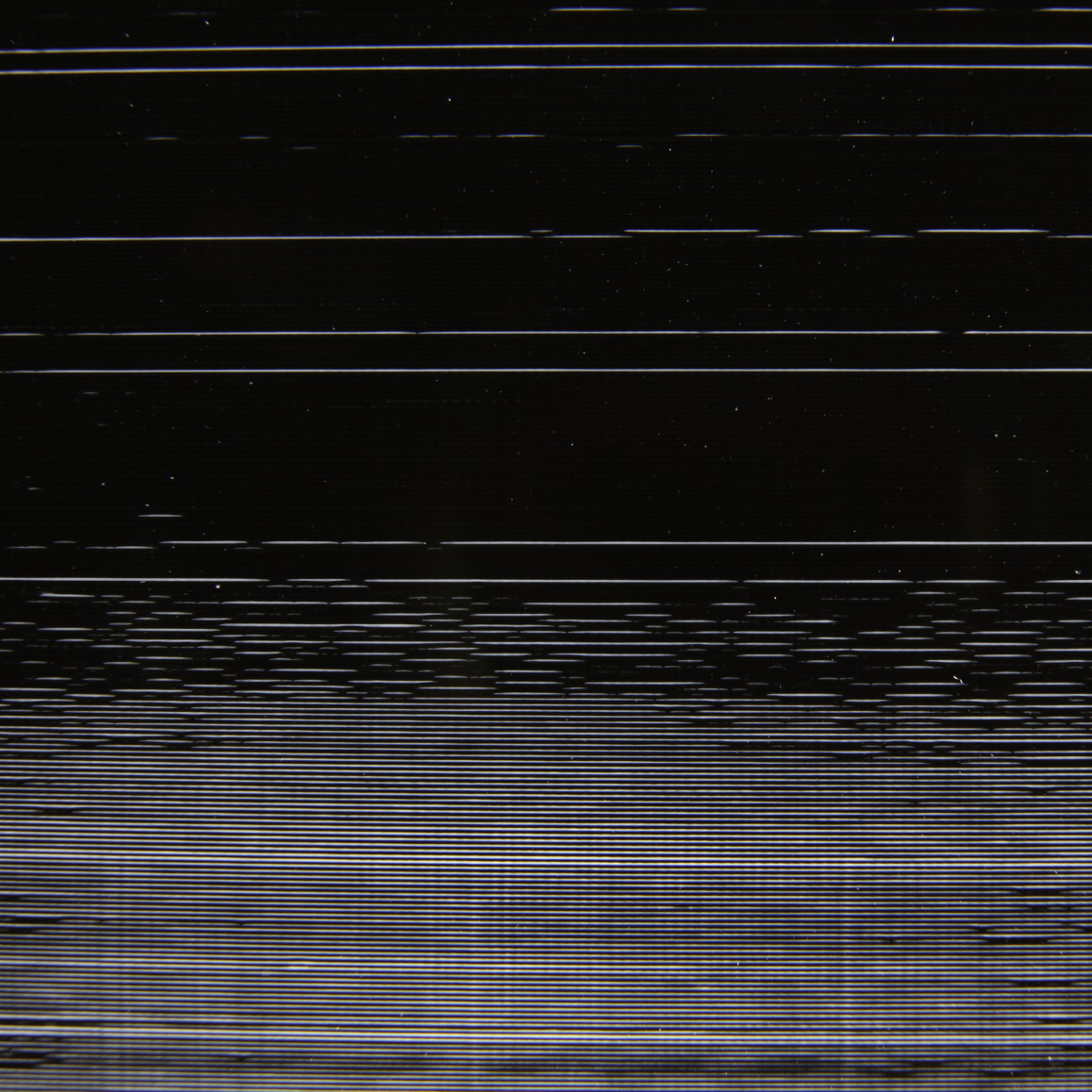}
        \caption{Fade}
	\label{fig:fade}
    \end{subfigure}
	\begin{subfigure}[b]{0.19\columnwidth}
        \centering
         \includegraphics[width=\textwidth]{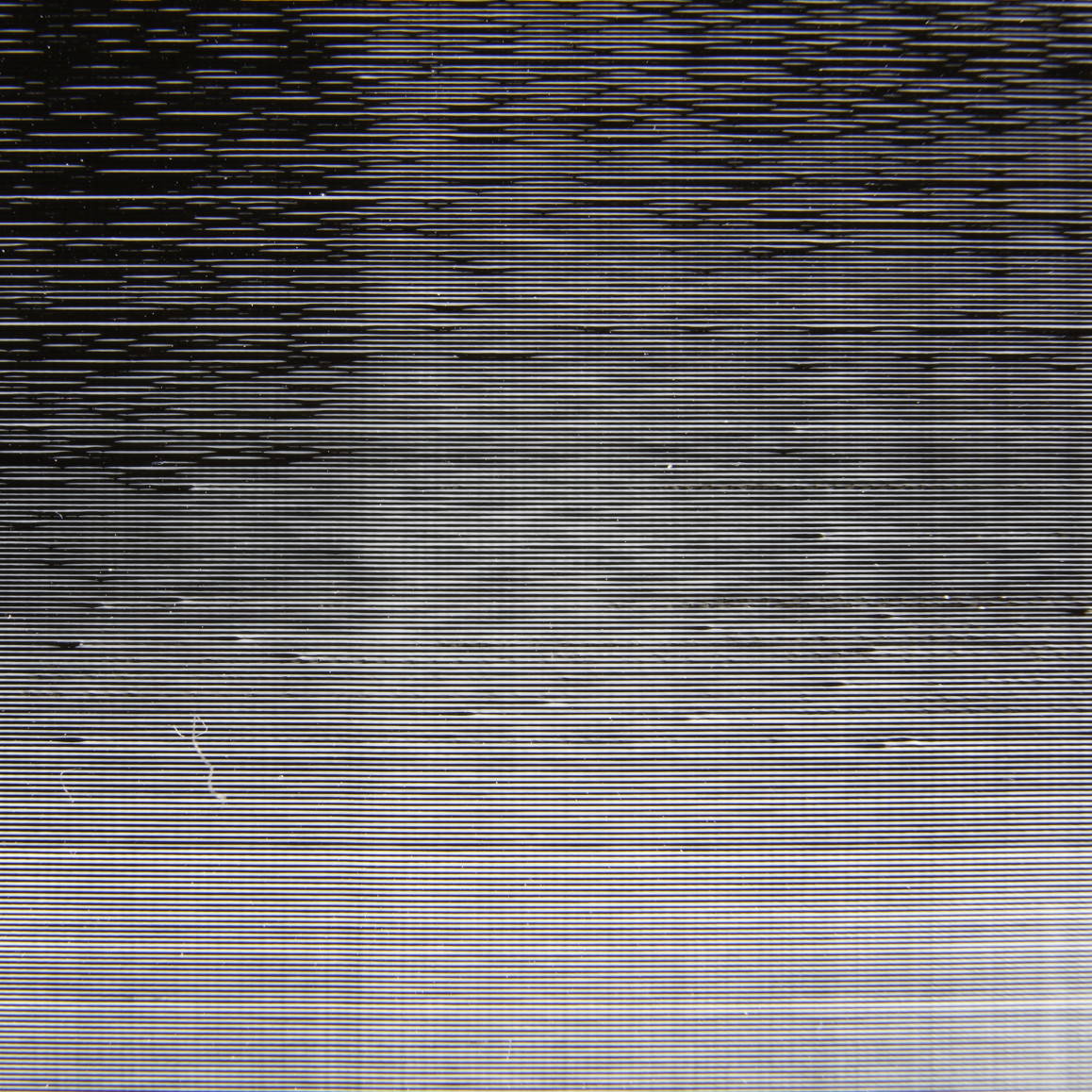}
        \caption{Swaying}
	\label{fig:swaying}
    \end{subfigure}
    \begin{subfigure}[b]{0.19\columnwidth}
        \centering
         \includegraphics[width=\textwidth]{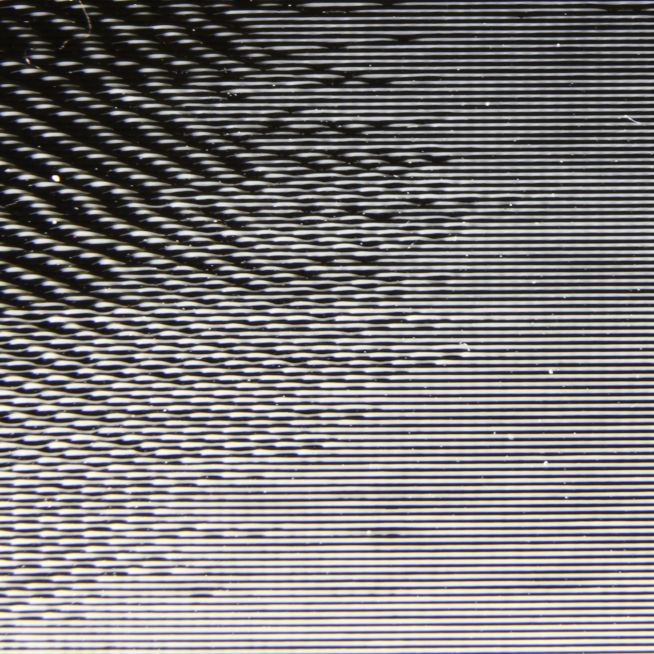}
        \caption{Ringing}
	\label{fig:ringing}
    \end{subfigure}
    \begin{subfigure}[b]{0.19\columnwidth}
        \centering
         \includegraphics[width=\textwidth]{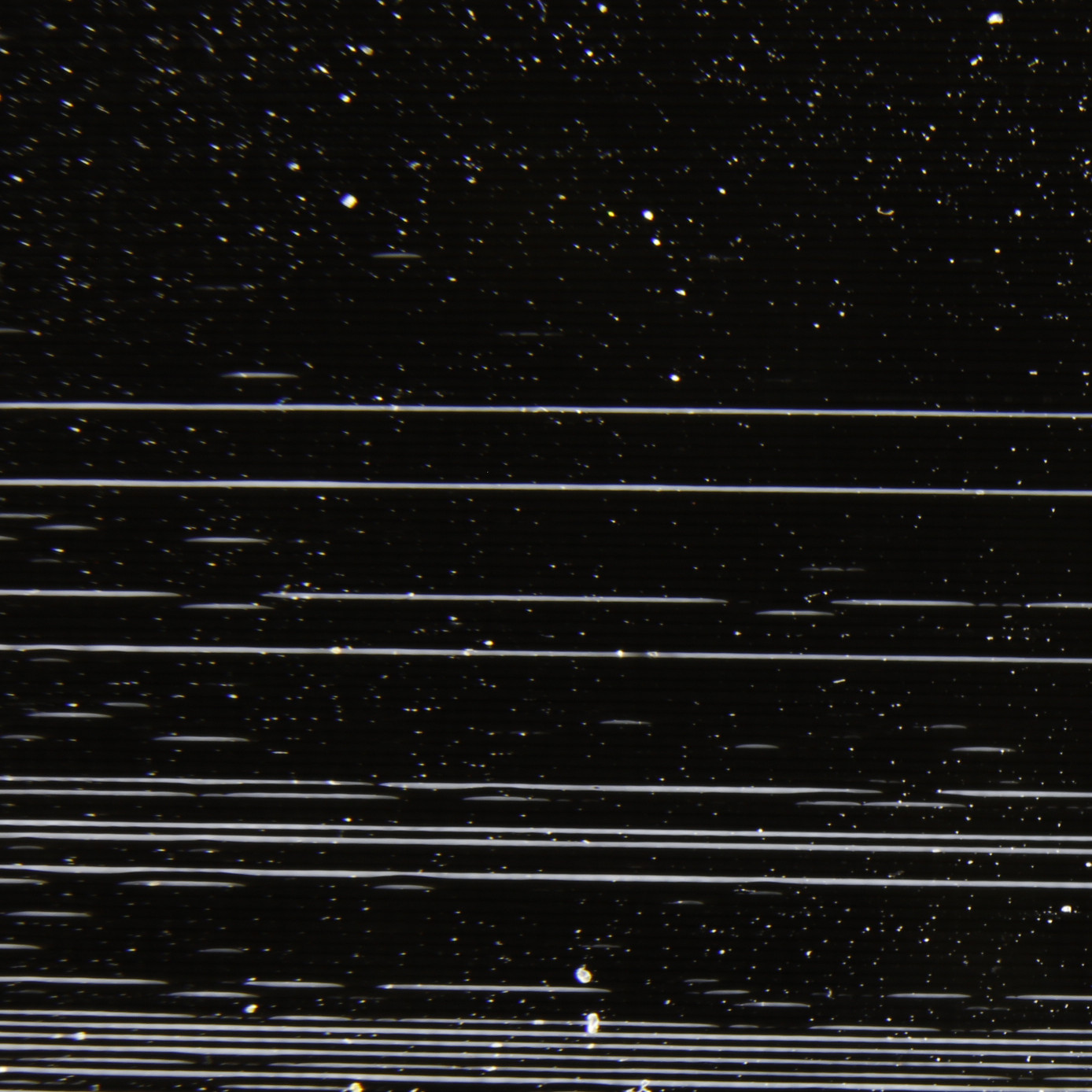}
        \caption{Banding}
	\label{fig:banding}
    \end{subfigure}
    \begin{subfigure}[b]{0.19\columnwidth}
        \centering
         \includegraphics[width=\textwidth]{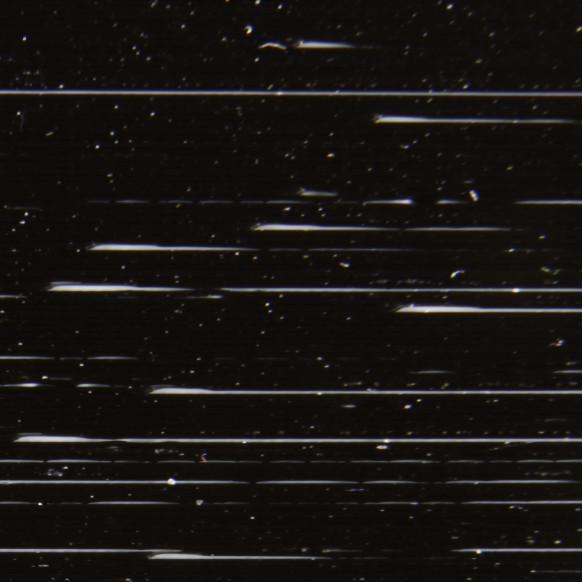}
        \caption{Blips}
	\label{fig:blips}
    \end{subfigure}
    \caption{Various artifacts in the FDM printing process.}
    \label{fig:inaccuracy}
\end{figure}

Figure~\ref{fig:inaccuracy} shows the artifacts that could be distinguished in our experimental results.
Note that there are potentially more types of defects, caused by different faulty elements in the FDM printing system and other variations in print parameters \cite{N.Turner2014}.
Vertical waves of lighter and darker colors, as can be seen in figure~\ref{fig:fade}, might be caused by inaccuracies in the bed positioning or by temperature gradients due to the heated build plate.
Slightly bent axles could account for low frequency horizontal wave patterns as can be seen in figure~\ref{fig:swaying}.
Jerks of the print head can cause ringing, which results in a high frequency horizontal wave pattern in the perceived luminance of vertical surfaces (see figure~\ref{fig:ringing}).
Banding - high frequency, vertical variations (see figure~\ref{fig:banding}), can be caused by various (combined) effects, that stem from inaccuracies in the printing system and/or variation in printing parameters. 
Small discontinuities in the surface, shown in figure~\ref{fig:blips} are caused by excess material oozing from the unused nozzle onto the model, while printing with the other nozzle. 
Inaccurate Z offsets between the two nozzles can lead to one of both filaments sagging more than the other, which can be seen in figure~\ref{fig:micro}, where the black layer is more compressed that the white layer.
Visual artifacts in horizontal hatching (used in figure~\ref{fig:lena_horizontal}) may result from reduced layer bonding when printing thinner lines.
Most of the inaccuracies described above, might be greatly reduced by using a single mixing nozzle, rather than an FDM system with two separate nozzles.

While these inaccuracies also affect the technique proposed by \citeauthor{reiner2014dual}, the resulting grayscale tone of their technique is less sensitive to these.
While the modulation of the amplitude of sine waves to the outer wall is limited to \SI{0.75}{\milli\meter} our hatching technique is limited to $1/2w=\SI{0.1}{\milli\meter}$.
This means that the technique presented in this paper is affected $7.5$ times as much by such inaccuracies.

One important observation from the sagging tests is that an overhang of \SI{0.2}{\milli\meter} already causes a full layer of vertical occlusion.
According to the geometrical sagging model this means that the area of the cross-section is increased by $ \pi h^2 - \pi (0.5 h)^2 - Cx . w . h \approx \SI{0.018}{\milli\meter\squared}$.
This means that sagging causes filament bleeding; extra filament is extruded, reducing the pressure in the nozzle, which impacts sagging in the subsequent region as well as influencing other aspects of the FDM process. 
This effect is not accounted for by our algorithm. 

Other visual artifacts in hatching stem from the fact that a grayscale value based variable offset is applied on a layer-by-layer basis.
The perceived grayscale value at a place in a given layer depends on the offsets applied at that place in the layer below and the layer above.
Visual artifacts occur where the texture tones or face angles in an area wildly differ between layers.
The implementation of vertical and diagonal hatching assume consecutive layers have the same layer thickness.
Where this assumption is violated visual defects occur as well.


\subsection{Viewing angle dependency}
Similar to the technique of \citeauthor{reiner2014dual}, the perceived tone is dependent on the viewing angle. 
%
%
However, as the surface produced with 3D hatching is less irregular than with the dithering technique of \citeauthor{reiner2014dual}, the perceived tone depends less on the azimuth of the viewing angle.
In other words, when shifting the viewing angle sideways, the perceived grayscale values show little change, as you are looking along the print layers.
The largest effect on the perceived grayscale is to be expected in the elevation viewing angle ($\phi$ in figure ~\ref{fig:response_functions}).
The relative offset between black and white consecutive layers and the sagging effect both lead to more occlusion with more extreme viewing angles. 

In the following part the predicted ratios of visible filament according to our sagging model are compared to printed results for the case corresponding to $n=\SI{0}{\degree}$ and $\phi=\alpha$ (see figure~\ref{fig:response_functions}).
18 vertical walls of $\SI{40}{}\times\SI{50}{\milli\meter}$ were printed, with a relative offset ranging from $\SI{-0.2}{\milli\meter}$ to $\SI{+0.2}{\milli\meter}$. 
Note that the artifacts described above, were also present in these printed samples to varying degrees, which influence the measurement result. 

\begin{figure}
    \centering
   \begin{subfigure}[t]{.49\columnwidth}
        \centering
         \includegraphics[height=.7\textwidth]{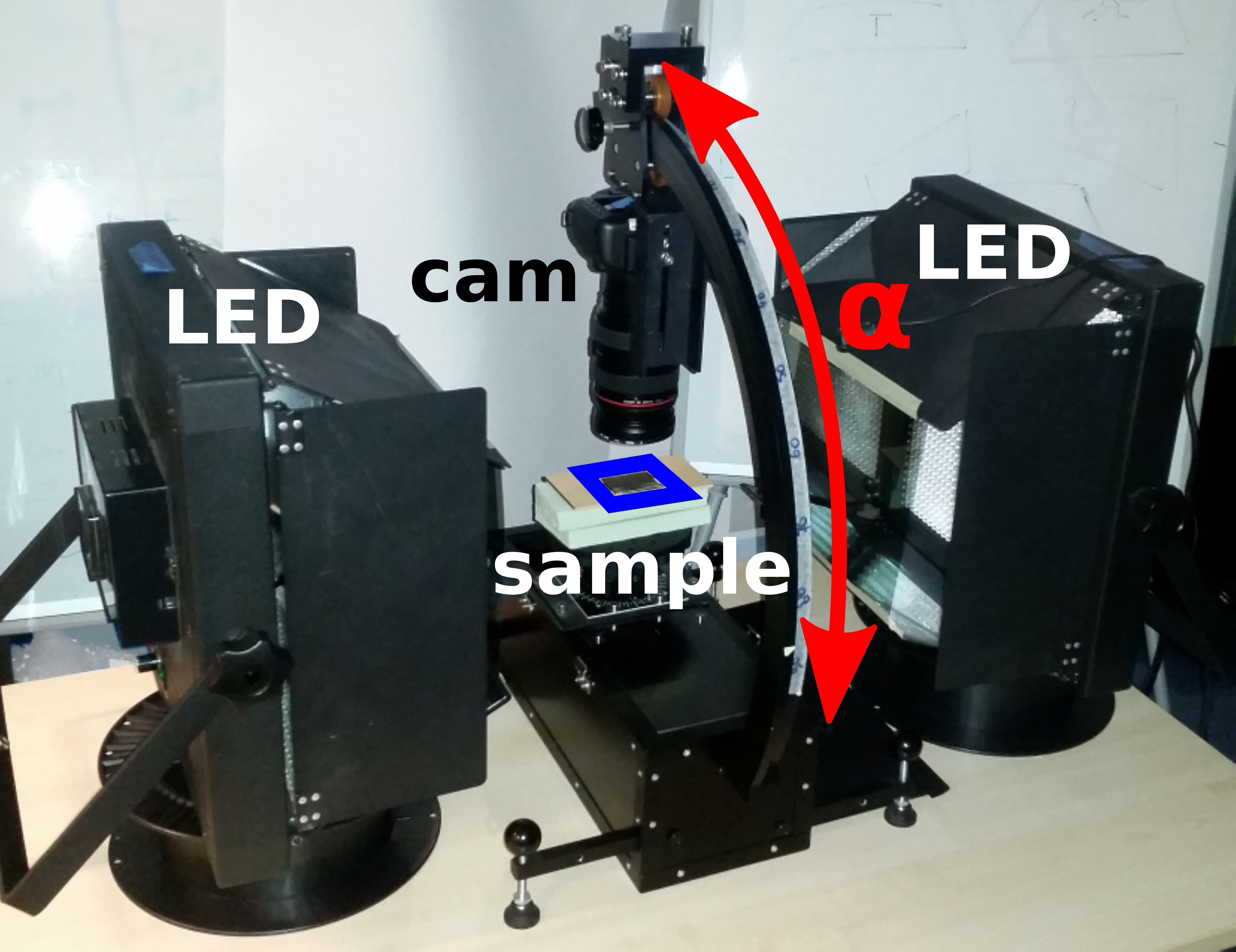}
        \caption{Gonioreflectometer setup}
	\label{fig:gonio_setup}
    \end{subfigure}
     \begin{subfigure}[t]{0.49\columnwidth}
         \centering
          \includegraphics[height=.7\textwidth]{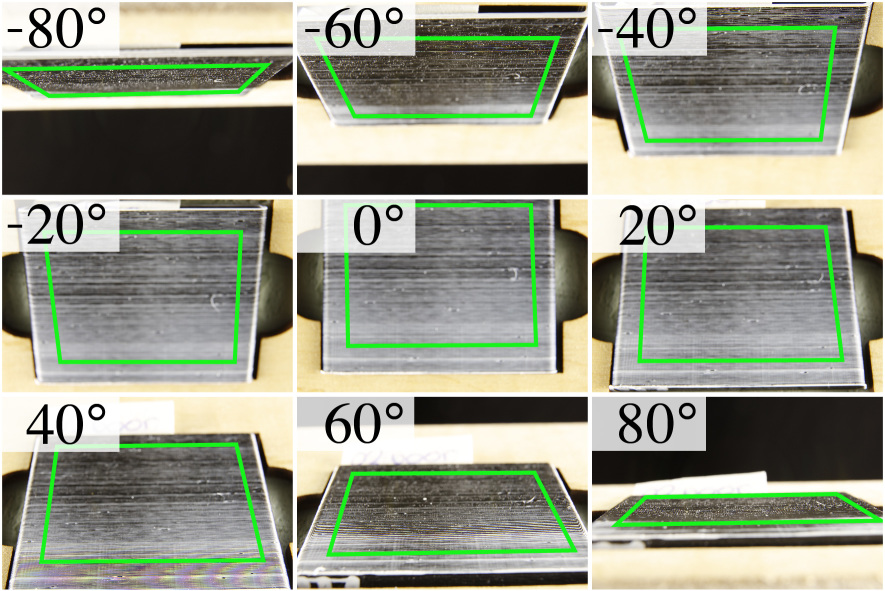}
         \caption{Sample at various viewing angles $\alpha$} 
 	\label{fig:gonio_samples}
     \end{subfigure}
%
%
    \caption{Measurement of printed samples}
    \label{fig:gonio_setup_and_samples}
\end{figure}

In order to capture these prints under various angles, a Canon 5DS camera was installed in a gonioreflectometer setup.
Samples were illuminated using two LED array light sources, placed at either side, in a cross-polarized setup, to avoid specular reflections as much as possible (see figure~\ref{fig:gonio_setup}) \cite{Hecht2001}.
The white balance of the camera was set using a completely white PLA print, and the exposure and aperture was set to avoid under- and over-exposure. 
Images were captured slightly out of focus to avoid moir\'e sampling effects, which are likely to occur due to the high frequency line patterns of the hatching technique. 

Samples were captured at viewing angles $\alpha$ ranging between \SI{-80}{\degree} and \SI{+80}{\degree} with \SI{10}{\degree} increments. 
A masked region of $\SI{35}{}\times\SI{40}{\milli\meter}$ was sampled in all images. Note that foreshortening occurs in the images due to the measurement angle (see figure ~\ref{fig:gonio_samples}). 
The luminance of the sampled area was recovered by converting the RGB image to gray-scale following the Luma standard \cite{poynton2004color}.
The mean luminance was calculated for every sample, applying the gamma compression of $2.2$ afterwards. 
The measured intensities are normalized relative to the minimum and maximum measured intensities.

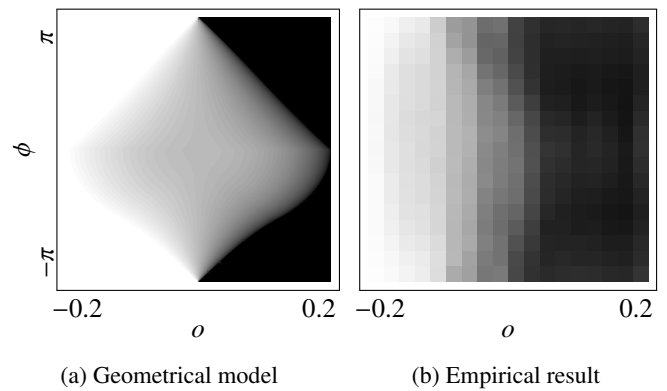
\begin{figure}[b]
\centering
    \begin{subfigure}[t]{.49\columnwidth}
    \centering
        \begin{tikzpicture}
            \node (img)  {  \fbox{\includegraphics[height=.8\textwidth]{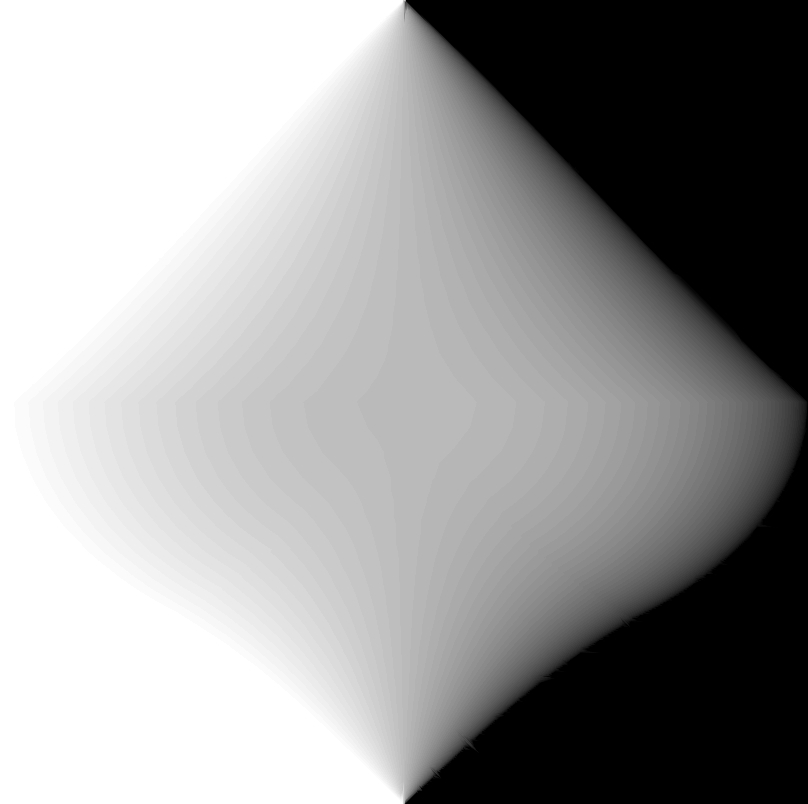}} };
            \node[left=of img, node distance=0cm, rotate=90, anchor=center,yshift=-0.7cm] {$\phi$};
            \node[below left=of img, node distance=0cm, rotate=90, anchor=center,yshift=-1.0cm,xshift=1.5cm] {$-\pi$};
            \node[above left=of img, node distance=0cm, rotate=90, anchor=center,yshift=-1.0cm,xshift=-1.5cm] {$\pi$};
            \node[below=of img, node distance=0cm, yshift=.8cm] {$o$};
            \node[below left=of img, node distance=0cm, anchor=center, yshift =.9cm, xshift=1.4cm] {$-0.2$};
            \node[below right=of img, node distance=0cm, anchor=center, yshift =.9cm, xshift=-1.4cm] {$0.2$};
        \end{tikzpicture}
    \caption{Geometrical model}  
    \label{fig:gonio_prediction}  
    \end{subfigure}
\centering
    \begin{subfigure}[t]{.49\columnwidth}
    \centering
        \begin{tikzpicture}
            \node (img)  {  \fbox{\includegraphics[height=.8\textwidth]{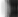}} };
            \node[below=of img, node distance=0cm, yshift=.8cm] {$o$};
            \node[below left=of img, node distance=0cm, anchor=center, yshift =.9cm, xshift=1.4cm] {$-0.2$};
            \node[below right=of img, node distance=0cm, anchor=center, yshift =.9cm, xshift=-1.4cm] {$0.2$};
        \end{tikzpicture}
    \caption{Empirical result}  
    \label{fig:gonio_results}  
    \end{subfigure}
    \caption{Luminance value for different amounts of overhang $o$ and viewing angles $\phi$ as predicted by our model (\subref{fig:gonio_prediction}) and as obtained from experimental results (\subref{fig:gonio_results}).}  
    \label{fig:gonio}  
\end{figure}


Figure~\ref{fig:gonio_results} show the measured intensities for different overhangs from different viewing positions.
Because of various visual artifacts and lighting conditions, comparing these results to figure~\ref{fig:gonio_prediction} doesn't yield the required insights to verify or falsify our geometrical model of sagging.
However, looking at the differences between the two graphs can shed some light on the influence of the lighting conditions at play.

These results do not show a 50-50 gray line for all viewing angles when there is no offset applied.
While a geometrical model should always satisfy such a condition, in reality less light is reflected to more extreme viewing angles than to a viewing angle perpendicular to the surface.

The fact that the graph in figure~\ref{fig:gonio_results} is not rotationally symmetric around $o=0$ can also not be explained by a geometrical model of sagging which is equally applied to both filaments.
Deviation from this symmetry could stem from multiple sources.
In the microscope image (figure~\ref{fig:micro}) the black filament appears to sag in a different way from white filament;
translucency and subsurface scattering might account for white filament appearing darker when enclosed by black filament;
shadows might play a role
and changing the shape of the outlines can contribute to different amounts of light being refracted in a specific direction.
Understanding the influence of these factors is needed to calibrate the appearance of full color FDM prints.


\section{Conclusion and future work}
We presented hatching halftoning techniques for 3D prints from dual extrusion FDM printing systems.
We show how to perform offsets of varying amplitude on polygonal outlines and specifically how to determine the amplitude of offsets required for achieving the right grayscale tones on the surface of 3D prints.
We investigate the phenomenon called \emph{sagging} and propose a model for it, which serves as basis of a thorough tone calibration, which is future work.
The results demonstrate the ability to manufacture objects with the appearance of full grayscale textures, while the techniques have little effect on printing time (maximum 20\% more than monochrome). 
Because the way in which layers are built up is unaltered, structural properties of the resulting product are unaffected.
The applicability of our hatching technique stands irrespective of the geometry and surface angle.
While it obtains a higher resolution than the technique presented by \citeauthor{reiner2014dual}, it is also more sensitive to inaccuracies in the mechanical FDM printing system.
This technique is applicable to any dual extrusion FDM system, although using a mixing nozzle can alleviate a lot of accuracy problems otherwise affecting the perceived grayscale tones.



One possible course of future research would be focused on improving on the unreliability caused by sagging.
A dithering technique might be adopted for near horizontal surfaces and surfaces with extremal grayscale tones.
The technique presented by \citeauthor{reiner2014dual} could serve this purpose once a way to gradually shift from their technique to the hatching technique for more horizontal surfaces.
The ability to apply the dithering technique gradually for more extreme combinations surface angle and texture color is required in order for the algorithm to be stable;
introducing a hard cutoff boundary between hatching and dithering introduces problems for surfaces around that cutoff boundary with slight variations in texture tone or surface angle.

A thorough tone calibration is needed to reproduce grayscale tones correctly.
In order to get a grip on the different factors which influence the perceived tone, different factors could be investigated,
for example material translucency, subsurface scattering, shadows induced by the offsetting and various lighting conditions.
Together with a verified geometrical model of sagging, the tone calibration results can then be explained.
The formula by which the print speed is computed from the line width in horizontal hatching also provides an opportunity for tone calibration.
Given that thin lines appear translucent, additional tone calibration might impact the required amount of material per line segment, which affects the line width modulation.

One could adopt a similar technique for FDM printers which have more than two extruders; if it is capable of printing with cyan, magenta, yellow, black and white filament, hatching could be used to produce full color prints.
When different or less colors are available, a mapping between color spaces should be performed to make the print appear as close to the textured mesh as possible.

A different line of research could be devoted to determining the optimal viewing angle from the model geometry and the texture image.
A more natural viewing angle could be obtained from the surface normal vector of the nearest point on the convex hull of the full geometry.
Where the texture image has more high frequency detail the optimal viewing angle could be determined more by the local surface angle of the model.
A deeper investigation of how humans look at 3D objects is needed to determine what the optimal viewing angle is at any point on the surface.

This paper presents techniques to print heterogeneous surface colors on FDM printing systems.
Future endeavors could be devoted to enabling other heterogeneous surface properties for FDM, such as surface roughness and specularity.
Going beyond just the surface of the mesh it could be researched how heterogeneous volumetric properties could be achieved in FDM.
What process planning techniques can be used to satisfy a heterogeneous infill density specification?

\section*{Acknowledgements}
We would like to thank Leo Haslam (Blockade figurine), Beerend Groot (tin can) and COMSOL (connecting rod) for permission to use their models.
We would also like to thank Jun Wu and Charlie Wang for suggestions and improvements.

Funding: This work was supported by Ultimaker.

\bibliographystyle{cag-num-names}
\bibliography{bibliography}

\end{document}